\def\ll_lsun{log$({L/\rm L_{\odot}})$~}  
  
\def\masa_msun{$M/ \rm M_{\odot}$~}  
  
\def\m_mstar{$M/M_{*}$~}

\def\v4334{\mbox{V4334 Sgr}}

\def\112{\mbox{SDSS J111215.82+111745.0}}  
\def\840{\mbox{SDSS J184037.78+642312.3}}  
\def\518{\mbox{SDSS J151826.68+065813.2}}  
\def\614{\mbox{SDSS J161431.28+191219.4}}  
\def\228{\mbox{SDSS J222859.93+362359.6}}
\def\738{\mbox{PSR J1738+0333}}
\def\J1618{\mbox{SDSS~J161831.69+385415.15}}

\documentclass[structabstract]{aa}
\usepackage{graphicx}  
\usepackage{txfonts}
\usepackage{subfigure}
\usepackage{natbib}
\bibpunct{(}{)}{;}{a}{}{,} % to follow the A&A style
\begin{document}  
\title{Pulsating low-mass white dwarfs in the frame of new evolutionary 
sequences} 
\subtitle{V. Asteroseismology of ELMV white dwarf stars}  

\author{Leila M. Calcaferro\inst{1,2}, 
        Alejandro H. C\'orsico\inst{1,2}, \and 
        Leandro G. Althaus\inst{1,2}}  
\offprints{L. M. Calcaferro} 

\institute{$^1$ Grupo  de Evoluci\'on  Estelar y  Pulsaciones,  Facultad de 
           Ciencias Astron\'omicas  y Geof\'{\i}sicas, Universidad
           Nacional de La Plata, Paseo del Bosque s/n, (1900) La
           Plata, Argentina\\   $^{2}$ Instituto de Astrof\'{\i}sica
           La Plata, CONICET-UNLP, Paseo  del Bosque s/n, (1900) La
           Plata,
           Argentina\\  \email{lcalcaferro,acorsico,althaus@fcaglp.unlp.edu.ar}}
\date{Received}  

\abstract{Many pulsating low-mass white-dwarf stars have been detected in the
last years in the field of our Galaxy. Given that some of them exhibit
multiperiodic variation of brightness, it is possible to probe their
interiors through asteroseismology.}{We aim to present a detailed
asteroseismological study of all the known low-mass variable white
dwarf stars based on a complete set of fully evolutionary models
representative of low-mass He-core white dwarf stars.}{We employed
adiabatic radial and nonradial pulsation  periods  for low-mass white
dwarf models with stellar masses ranging from $0.1554$ to $0.4352\
M_{\sun}$, that were derived by simulating the nonconservative
evolution of a binary system consisting of an initially $ 1 M_{\sun}$
ZAMS star and a $1.4  M_{\sun}$ neutron star companion. We estimated
the mean period spacing for the stars under study (in the cases where
it was possible) and then we used the comparison between the observed
period spacing with the average of the computed period spacings for
our grid of models to constrain the stellar mass. We also employed the
individual observed periods of every known pulsating low-mass white
dwarf star, to search for a representative seismological model.}{We
found that even though the stars under analysis exhibit few periods
and the period fits show multiplicity of solutions, it is possible to
find seismological models whose mass and effective temperature are in
agreement with the values given by spectroscopy, for most of the
cases.  Unfortunately, we were not able to constrain the stellar
masses by employing the observed period spacing because, in general,
the periods exhibited by these stars are very few. In the two cases
where we could extract the period spacing from the set of observed
periods, this method led to values of the stellar masses substantially
larger than expected for this type of stars.}{The results presented in
this work show in the one hand, the need for further photometric
searches, and on the other hand, some improvements of the theoretical models,
in order to place the asteroseismological results on a firmer ground.} 
\titlerunning{Asteroseismology of ELMV WDs}  
\maketitle  
\authorrunning{Calcaferro et al.}  

%----------------------------------------------------------------  
   
\section{Introduction}  
\label{intro}  

White dwarf (WD) stars are the last stage in the life of the majority
of stars \citep{2008ARA&A..46..157W,2008PASP..120.1043F,review}.  Most
WDs have envelopes rich in H, and they define the spectral class DA WD
whose distribution peaks at $0.59 M_{\sun}$. It also shows a peak at
low mass: $M_{\star}/M_{\sun}  \lesssim  0.45$. These stars are
thought to be the result of strong  mass-loss episodes in interactive
binary systems, before the He flash during the red giant branch phase
of low-mass  stars
(\cite{2013A&A...557A..19A,2016A&A...595A..35I}, for recent studies).
At variance with average WDs with C and O cores, they are expected to
contain He cores, since He burning is avoided.  Specifically, this
interactive binary evolutionary scenario is thought to be the  most
plausible origin for  the so-called  extremely low-mass (ELM) WDs,
which  have masses below $\sim  0.18-0.20 M_{\sun}$.

In the last years, numerous low-mass WDs, including ELM WDs, have been
discovered via the ELM survey and the SPY and WASP  surveys
\citep[see][]{2009A&A...505..441K,  2010ApJ...723.1072B,
  2012ApJ...744..142B, 2011MNRAS.418.1156M, 2011ApJ...727....3K,
  2012ApJ...751..141K, 2013ApJ...769...66B, 2014ApJ...794...35G,
  2015MNRAS.446L..26K, 2015ApJ...812..167G}. The detection of
  pulsation $g$ (gravity) modes in some of
  them   \citep{2012ApJ...750L..28H,
  2013ApJ...765..102H,2013MNRAS.436.3573H,
  2015MNRAS.446L..26K,2015ASPC..493..217B,2017ApJ...835..180B} has
  given rise to a new class of variable white dwarfs, the ELMVs.
  These pulsating low-mass WDs provide us an exceptional chance for
  probing the interiors of  these stars and eventually to test their
  formation scenarios by employing the tools of
  asteroseismology. Since $g$ modes  in ELMVs are restricted mainly to
  the core regions \citep{2010ApJ...718..441S, 2012A&A...547A..96C,
  2014A&A...569A.106C}, we would be able to constrain  their core
  chemical structure.  Furthermore, as shown by  stability
  computations \citep{2012A&A...547A..96C,2013ApJ...762...57V,2016A&A...585A...1C},
  a combination of the $\kappa-\gamma$
  mechanism \citep{1989nos..book.....U} and the ``convective driving''
  mechanism \citep{1991MNRAS.251..673B}, both acting at the
  H-ionization region, excite long-period $g$ modes in agreement with
  observations. Moreover, some unstable short-period $g$  modes  could
  be  driven   by the $\varepsilon$ mechanism due to stable H
  burning \citep{2014ApJ...793L..17C}.

In addition to ELM stars, there are several objects considered as their precursors,
the so-called pre-ELMs. These stars exhibit metals in their atmospheres
\citep[e.g.][]{2014ApJ...781..104G,2014MNRAS.444.1674H,2016A&A...595A..35I}. Interestingly
enough, pulsations in a number of objects have been detected in the last years
\citep{2013Natur.498..463M, 2014MNRAS.444..208M,2016ApJ...821L..32Z,
  2016ApJ...822L..27G,2016A&A...587L...5C}. Evolutionary models that consider only element
  diffusion cannot explain these properties 
\citep[e.g.][]{2016A&A...588A..74C,2016A&A...595L..12I} and might be an indication that
the missing physics could
impact also the evolution of the objects on the cooling track (e.g., the thickness of the
H envelope). Moreover, there are indications that a pre-ELMV WD will be later
observed as an ELMV \citep{2017ASPC..509..347F}.

The definition of an ELM WD is still under debate. In the context of the ELM
survey, an ELM WD is defined as an object with surface gravity of
$5 \lesssim \log(g) \lesssim 7$ and effective temperature in the range of
$8000 \lesssim T_{\rm eff} \lesssim 22\,000\ $
\citep[e.g.][]{2010ApJ...723.1072B,2011ApJ...727....3K,2016ApJ...818..155B}.
In addition, an ELM WD should
be part of a tight binary system, and therefore show short-period or high-amplitude
velocity variability \citep[e.g.][]{2017ApJ...839...23B}. \cite{2014A&A...569A.106C}
suggests to define an
ELM WD as a WD that does not undergo H shell flashes as the pulsational
properties are quite different as compared with the systems that experience flashes.
However, this mass limit depends on the metallicity of the progenitor stars
\citep{2016A&A...595A..35I}.

WD asteroseismology has already proven to be a very useful technique
for peering into the interior of several pulsating WDs, and it has
been applied by employing two different methodologies: one considering
stellar models with parametrized chemical composition profiles, and
another involving fully evolutionary models characterized by a
consistent chemical structure. The former has the advantage of
allowing a full exploration of the parameter space (the
total mass, the mass of the H envelope, the chemical composition of
the core, among others) to find an  optimal asteroseismological
model. Examples of this approach are the pioneer works
by \cite{1998ApJS..116..307B,
2001ApJ...552..326B}. More recent works using this avenue are
from \cite{2006A&A...446..223P,2006A&A...453..219P,2008ApJ...675.1505B,
2008MNRAS.385..430C,2009MNRAS.396.1709C,2013MNRAS.432..598P,2016MNRAS.461.4059B}
and the recent developments of the core parameterization by \cite{2016ApJS..223...10G,
2017A&A...598A.109G,2017ApJ...834..136G}. The second approach,
developed at La Plata Observatory, is different but complementary
as it employs fully evolutionary models which are the result of the
complete evolution of the progenitor stars, from the Zero Age Main
Sequence (ZAMS) until the WD phase. Examples of the application of
this method to GW Virginis stars (pulsating PG1159 stars) are the
works by \cite{2007A&A...461.1095C,2007A&A...475..619C,
2008A&A...478..869C,2009A&A...499..257C,2014MNRAS.442.2278K} and
\cite{2016A&A...589A..40C}. Also, it has been applied in DBV WDs
(He-rich atmosphere)
by \citet{2012A&A...541A..42C,2014A&A...570A.116B}.  Regarding ZZ Ceti
stars, this approach has been successfully employed
by \cite{2012ApJ...757..177K,2012MNRAS.420.1462R,2013ApJ...779...58R}.
In particular, this method has the value added that the chemical
structure of the background models is consistent with the pre-WD
evolution.

\begin{figure}
\begin{center}
\includegraphics[clip,width=9 cm]{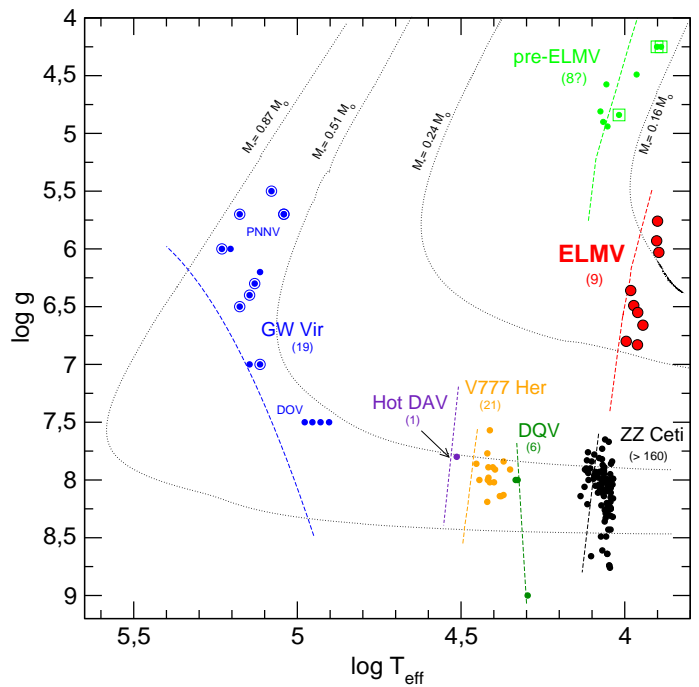} 
\caption{The location of the known ELMVs (red circles)
along with the other several classes of pulsating WD stars (dots of
different colors) in the $\log T_{\rm eff}-\log g$ plane.
The three stars emphasized with squares surrounding the light green
circles can be identified as pre-ELMV stars as well as SX Phe
and/or $\delta$ Scuti stars. In parenthesis we include the
number of known members of each class. Two post-VLTP (Very Late Thermal
Pulse) evolutionary tracks for H-deficient WDs and two evolutionary
tracks for low-mass He-core WDs are plotted for reference. Dashed
lines indicate the theoretical blue edge for the different
classes of pulsating WDs.}
\label{HR1} 
\end{center}
\end{figure}

Here, we apply our asteroseismological approach for the first time to
the complete set of the known ELMVs, whose spectroscopic parameters we
describe in Table~\ref{tablacero}, and whose list of periods
exhibited can be found in Tables~\ref{table1} to~\ref{table9}, along
with the corresponding frequencies and amplitudes.
\840 (hereafter J1840) is the first ELMV
discovered \citep{2012ApJ...750L..28H,2013MNRAS.436.3573H}. \112 (hereafter
J1112) was reported by \cite{2013ApJ...765..102H}. This
case is particularly interesting because this star shows seven
periods, and two of them are very short, probably associated with $p$
or radial modes. 
\518 (hereafter J1518) is the hottest ELMV hitherto, according to
\cite{2013ApJ...765..102H}. This star shows seven independent periods,
making possible a more detailed asteroseismological analysis. 
\614 (hereafter J1614)
is an ELMV, according to \cite{2013MNRAS.436.3573H}. \228 (hereafter J2228)
is currently the coolest EMLV, according to \cite{2013MNRAS.436.3573H}.
This star exhibits only three independent periods in
the range $\sim [3255-6235]$\ s, so these periods seem to be
approximately in the asymptotic regime \citep[see][]{2014A&A...569A.106C}.
In particular, the period $6234.9\ $s is the
longest period ever measured in a pulsating WD star.
\738, is a millisecond pulsar that has an ELMV companion (which we shall call,
for short, J1738), according to \cite{2015MNRAS.446L..26K}. This case is
particularly interesting because it is the only binary system with a
millisecond pulsar and a pulsating WD.
\J1618 (hereafter J1618) is an ELMV, according
to \cite{2015ASPC..493..217B}. Finally, SDSS J1735+2134 (hereafter
J1735) and SDSS J2139+2227 (hereafter J2139) are two recently detected
ELMVs, according to \cite{2017ApJ...835..180B}. In particular, the former
has very long periods, which seem to be in the asymptotic regime of nonradial
$g$ modes \citep[see][]{2014A&A...569A.106C}. It is worth mentioning
how the spectroscopic masses of the ELM WDs are determined. \cite{2017ApJ...839...23B}
showed that for the same metallicity of the progenitor stars there is a 15\% difference
in the mass of ELM WDs for the same $\log(g)$ and $T_{\rm eff}$ parameters, using either the
\cite{2013A&A...557A..19A} evolutionary tracks or the \cite{2016A&A...595A..35I}
evolutionary tracks. We can adopt this difference as the true uncertainty in the
spectroscopic determination of the masses of ELM WDs.
In Fig. \ref{HR1} we show the location of
the different families of pulsating WDs, including all the known ELMV
stars (red circles).  The total number of ELMVs rises to nine, because
there is a high probability that the star discovered
by \cite{2017ApJ...835..180B}, SDSS J1355+1956, is a $\delta$ Scuti
pulsator, as claimed by these authors. In this sense, it is important
to stress here that some of the stars under analysis in this work may
not be pulsating ELM WDs. The analysis done by \cite{2017ApJ...839...23B}
suggests that
there are actually only four pulsating ELM WDs: J1840, J1112, J1518 and J1738. In addition,
as discussed by \cite{2015ASPC..493..217B,2017ApJ...835..180B}, the stars J1618, J1735 and
J2139 may not be ELM WD stars. A measurement of the rate of period change for
these stars could help to shed light on this
issue \citep{2017A&A...600A..73C}. Despite of this, given the
exploratory nature of this work, we will consider that these stars are
genuine ELMVs and they will be included in our analysis.

\begin{table*}[t]
\centering
\caption{Stellar  parameters (derived using 1D and 3D model atmospheres)
and observed pulsation properties of all the known ELMV WD stars.}
\begin{tabular}{cccccccc}
\hline 
\hline
\vspace{1mm}
Star & $T^{1D}_{\rm eff}$ & $\log(g)^{1D}$ & $M^{(1D)}_{\star}$ & $T^{3D}_{\rm eff}$ &  $\log(g)^{3D}$ & $M^{(3D)}_{\star}$ & Period range \\
       &  [K]            & [cgs]    &    [$M_{\sun}$] &   [K]          & [cgs]    &    [$M_{\sun}$] &  [s] \\
\hline
J1840  & $9390 \pm 140$ &  $6.49 \pm 0.06$  & 0.183$^{\rm a,b}$ & $9120 \pm 140$ & $6.34 \pm 0.05$  & 0.177$^{\rm c}$ & [1164-4445] \\ 
J1112  & $9590 \pm 140$ &  $6.36 \pm 0.06$  & 0.179$^{\rm d}$  & $9240 \pm 140$ & $6.17 \pm 0.06$  & 0.169$^{\rm c}$ & [108-2856] \\ 
J1518  & $9900 \pm 140$ &  $6.80 \pm 0.05$  & 0.220$^{\rm d}$  & $9650 \pm 140$ & $6.68 \pm 0.05$  & 0.197$^{\rm c}$ & [1335-3848] \\ 
J1614  & $8800 \pm 170$ &  $6.66 \pm 0.14$  & 0.192$^{\rm b}$  & $8700 \pm 170$ & $6.32 \pm 0.13$  & 0.172$^{\rm c}$ & [1184-1263] \\
J2228  & $7870 \pm 120$ &  $6.03 \pm 0.08$  & 0.152$^{\rm b}$  & $7890 \pm 120$ & $5.78 \pm 0.08$  & 0.142$^{\rm c}$ & [3255-6235] \\
J1738  & $9130 \pm 140$ &  $6.55 \pm 0.06$  & 0.181$^{\rm e}$  & $8910 \pm 150$ & $6.30 \pm 0.10$  & 0.172$^{\rm c}$ & [1788-3057] \\
J1618  & $9144 \pm 120$ &  $6.83 \pm 0.14$  & 0.220$^{\rm f}$  & $8965 \pm 120$ & $6.54 \pm 0.14$  & 0.179$^{\rm c}$ & [2543-6126] \\
J1735  & ---            & ---               &   ---          & $7940 \pm 130$  & $5.76 \pm 0.08$ & 0.142$^{\rm g}$ & [3363-4961] \\
J2139  & ---            & ---               &   ---          & $7990 \pm 130$  & $5.93 \pm 0.12$ & 0.149$^{\rm g}$ & [2119-3303] \\
\hline
\hline
\end{tabular} 
\label{tablacero}

{\footnotesize  Notes: 
$^{\rm a}$\citet{2012ApJ...750L..28H}.
$^{\rm b}$\citet{2013MNRAS.436.3573H}.
$^{\rm c}$ Determined using the corrections for 3D effects by \citet{2015ApJ...809..148T}.}
$^{\rm d}$\citet{2013ApJ...765..102H}.
$^{\rm e}$\citet{2015MNRAS.446L..26K}.  
$^{\rm f}$\citet{2015ASPC..493..217B}.  
$^{\rm g}$\citet{2017ApJ...835..180B}.
\end{table*}

In this paper we report a further step in the study of low-mass WD stars by
performing an asteroseismological analysis of all the known ELMVs.
This is the fifth work of a series dedicated to these stars. The first one
\citep{2014A&A...569A.106C} was focused on the adiabatic properties
of these stars; the second one \citep{2016A&A...585A...1C} was dedicated to
the nonadiabatic pulsation stability features of these stars. The third work
\citep{2016A&A...588A..74C} was aimed at studying the pulsation properties of 
the pre-ELMV WDs. The fourth paper \citep{2017A&A...600A..73C} was focused on
studying the theoretical temporal rates of period change of ELMV and pre-ELMV stars.
In this work, we follow the approach that employs fully evolutionary models resulting
from the complete evolution of the progenitor stars. The employment of
fully evolutionary models is a crucial requirement because some models
(particularly those with the lowest mass) are characterized by strong
H-nuclear burning that depends sensitively on the thickness of the H
envelope, a quantity that results from the previous evolution. We
employ the adiabatic radial ($\ell= 0$) and non-radial ($\ell= 1,2$)
$p-$ and $g-$mode pulsation periods computed
in \cite{2014A&A...569A.106C} on low-mass He-core WD evolutionary
models with stellar masses  ranging from $0.1554$  to $0.4352  \
M_{\sun}$, extracted from the computations
of \cite{2013A&A...557A..19A}, that take into account the binary
evolution of the progenitor  stars.

The paper is organized as follows. A brief summary of the stellar
models and the pulsational code employed is provided in
Sect. \ref{evolutionary}.  In Sect. \ref{analysis} we describe the
asteroseismological analyses carried out. Next, in Sect. \ref{estimation},
we try to determine (when possible) the observed period spacing for
the target stars, and then in Sect. \ref{psp} we constrain the stellar
mass by comparing the observed period spacing with the average of
the computed period spacings. In Sect. \ref{fitting}, we search for
the best-fit asteroseismological model by comparing the individual
periods from each ELMV star with theoretical periods from our grid of
models.  Finally, in Sect. \ref{conclusions} we summarize the main
findings of this work.

\begin{table}[ht]
\centering
\caption{List of the 5 independent frequencies in the data 
of J1840 from \cite{2012ApJ...750L..28H}.}
\begin{tabular}{cccccccccccc}
\hline
\hline  
$\Pi$ [s] & Freq.[$\mu$Hz] & Ampl. [mmag] \\
\hline  
$1164.15 \pm 0.38 $ & $859.0 \pm 0.29$    &  $1.78 \pm 0.29$\\
$1578.7 \pm 0.65 $  & $633.43 \pm 0.26$   &  $2.831 \pm 0.41$\\
$2376.07 \pm 0.74$     & $420.86 \pm 0.13$   &  $4.817 \pm 0.46$\\
$3930.0 \pm 300 $     & $254.0 \pm 19$      &  $2.7 \pm 2.0$\\
$4445.3 \pm 2.4$    & $224.96 \pm 0.12$   &  $7.6 \pm 1.6$\\
\hline
\hline  
\end{tabular}
\label{table1}
\end{table} 

\begin{table}[ht]
\centering
\caption{List of the 7 independent frequencies in the data 
of J1112 from \cite{2013ApJ...765..102H}.}
\begin{tabular}{cccccccccccc}   
\hline
\hline  
$\Pi$ [s] & Freq.[$\mu$Hz] & Ampl. [mmag] \\
\hline  
$107.56   \pm 0.04$  & $9297.4 \pm 3.6$      & $0.38 \pm 0.14$\\
$134.275  \pm 0.001$ & $7447.388 \pm 0.0100$ & $0.44 \pm 0.08$\\
$1792.905 \pm 0.005$ & $557.7542 \pm 0.0017$ & $3.31 \pm 0.08$\\
$1884.599 \pm 0.004$ & $530.6170 \pm 0.0011$ & $4.73 \pm 0.08$\\
$2258.528 \pm 0.003$ & $442.7662 \pm 0.0007$ & $7.49 \pm 0.08$\\
$2539.695 \pm 0.005$ & $393.7480 \pm 0.0007$ & $6.77 \pm 0.09$\\
$2855.728 \pm 0.010$ & $350.1734 \pm 0.0013$ & $3.63 \pm 0.09$\\
\hline
\hline  
\end{tabular}
\label{table2}
\end{table} 

\begin{table}[ht]
\centering
\caption{List of the 7 independent frequencies in the data 
of J1518 from \cite{2013ApJ...765..102H}.}
\begin{tabular}{cccccccccccc}
\hline
\hline  
$\Pi$ [s] & Freq.[$\mu$Hz] & Ampl. [mmag] \\
\hline  
$1335,318 \pm 0.003$ & $748.8855 \pm 0.0015$ &  $13.6 \pm 0.6$\\ 
$1956,361 \pm 0.003$ & $511.1532 \pm 0.0007$ &  $18.1 \pm 0.3$\\
$2134,027 \pm 0.004$ & $468.5976 \pm 0.0008$ &  $14.2 \pm 0.4$\\
$2268,203 \pm 0.004$ & $440.8777 \pm 0.0007$ &  $21.6 \pm 0.2$\\
$2714,306 \pm 0.003$ & $368.4183 \pm 0.0005$ &  $21.6 \pm 0.9$\\
$2799.087 \pm 0.005$ & $357.2593 \pm 0.0007$ &  $35.4 \pm 0.6$\\
$3848.201 \pm 0.009$ & $259.8617 \pm 0.0006$ &  $15.7 \pm 0.3$\\
\hline
\hline  
\end{tabular}
\label{table3}
\end{table}

\begin{table}[ht]
\centering
\caption{List of the 2 independent frequencies in the data 
of J1614 from \cite{2013MNRAS.436.3573H}.}
\begin{tabular}{cccccccccccc}
\hline
\hline  
$\Pi$ [s] & Freq.[$\mu$Hz] & Ampl. [mmag] \\
\hline  
$1184.106 \pm 0.064$ & $844.519 \pm 0.045$ &  $3.20\pm 0.10$\\
$1262.668 \pm 0.041$ & $791.974 \pm 0.026$ &  $5.94 \pm 0.11$\\
\hline
\hline  
\end{tabular}
\label{table4}
\end{table} 

\begin{table}[ht]
\centering
\caption{List of the 3 independent frequencies in the data 
of J2228 from \cite{2013MNRAS.436.3573H}.}
\begin{tabular}{cccccccccccc}
\hline
\hline  
$\Pi$ [s] & Freq.[$\mu$Hz] & Ampl. [mmag] \\
\hline  
$3254.5 \pm 2.1$ & $307.27 \pm 0.20$ & $2.34 \pm 0.14$\\
$4178.3 \pm 2.8$ & $239.33 \pm 0.16$ & $6.26 \pm 0.14$\\
$6234.9 \pm 6.0$ & $160.39 \pm 0.15$ & $1.94 \pm 0.23$\\
\hline
\hline  
\end{tabular}
\label{table5}
\end{table} 

\begin{table}[ht]
\centering
\caption{List of the 3 independent frequencies in the data 
of J1738 from \cite{2015MNRAS.446L..26K}.}
\begin{tabular}{cccccccccccc}  
\hline
\hline  
$\Pi$ [s] & Freq.[$\mu$Hz] & Ampl. [mmag] \\
\hline  
$1788 \pm 33$ & $559 \pm 10$ & $1.27 \pm 0.47$\\
$2656 \pm 80$ & $376 \pm 11$ & $1.15 \pm 0.47$\\
$3057 \pm 99$ & $327 \pm 11$ & $1.22 \pm 0.47$\\
\hline
\hline  
\end{tabular}
\label{table6}
\end{table} 

\begin{table}[ht]
\centering
\caption{List of the 3 independent frequencies in the data 
of J1618 from \cite{2015ASPC..493..217B}.}
\begin{tabular}{cccccccccccc}
\hline
\hline  
$\Pi$ [s] & Freq.[$\mu$Hz] & Ampl. [mmag] \\
\hline  
$2543.0 \pm 10$ & $393.2 \pm 1.6$ & $16 \pm 3$\\
$4935.21 \pm 0.07$ & $202.605 \pm 0.003$ & $56.3 \pm 1.3$\\
$6125.9 \pm 0.2$ & $163.240 \pm 0.006$ & $25.5 \pm 1.4$\\
\hline
\hline  
\end{tabular}
\label{table7}
\end{table} 

\begin{table}[ht]
\centering
\caption{List of the 4 independent frequencies in the data 
of J1735 from \cite{2017ApJ...835..180B}.}
\begin{tabular}{cccccccccccc}
\hline
\hline  
$\Pi$ [s] & Freq.[$\mu$Hz] & Ampl. [mmag] \\
\hline  
$3362.76 \pm 0.54 $   & $297.38 \pm 0.05$   &  $2.04 \pm 0.11$\\
$3834.54 \pm 0.42$    & $260.79 \pm 0.03$   &  $3.64 \pm 0.11$\\
$4541.88 \pm 0.24 $   & $220.172 \pm 0.013$ &  $7.60 \pm 0.11$\\
$4961.22 \pm 0.72$    & $201.56 \pm 0.03$   &  $3.38 \pm 0.11$\\
\hline
\hline  
\end{tabular}
\label{table8}
\end{table}

\begin{table}[ht]
\centering
\caption{List of the 3 independent frequencies in the data 
of J2139 from \cite{2017ApJ...835..180B}.}
\begin{tabular}{cccccccccccc}
\hline
\hline  
$\Pi$ [s] & Freq.[$\mu$Hz] & Ampl. [mmag] \\
\hline  
$2119.44 \pm 0.24 $    & $471.82 \pm 0.06$   &  $1.52 \pm 0.08$\\
$2482.32 \pm 0.54 $    & $402.85 \pm 0.09$   &  $1.02 \pm 0.08$\\
$3303.30 \pm 0.96 $    & $302.73 \pm 0.09$   &  $0.99 \pm 0.08$\\
\hline
\hline  
\end{tabular}
\label{table9}
\end{table}

\section{Evolutionary models and pulsational code}  
\label{evolutionary}  
  
In this work, we have employed the fully evolutionary models of
low-mass He-core WDs generated with  the {\tt LPCODE} stellar
evolution code.  This code computes in detail the complete
evolutionary stages which lead to the WD formation, allowing the study
of the WD evolution consistently  with the predictions of the
evolutionary  history of progenitors.  Details of {\tt LPCODE} can be
found in \citet{2005A&A...435..631A,
2009A&A...502..207A,2013A&A...557A..19A,2015A&A...576A...9A} and
references therein. Here, we briefly mention  the   ingredients
employed  which  are  relevant for  our analysis  of  low-mass,
He-core WD \citep[see][for details]{2013A&A...557A..19A}.  The
standard  Mixing Length  Theory (MLT)   for convection in the ML2
prescription is used
\citep[see][for its definition]{1990ApJS...72..335T}, however adiabatic
periods do not sensitively depend on the specific version of the MLT
theory of convection employed
\citep{1998ApJS..116..307B}.
We assumed the metallicity of the progenitor stars to be $Z = 0.01$.
We considered  the radiative opacities for arbitrary metallicity in
the range of 0 to  0.1 from the OPAL
project \citep{1996ApJ...464..943I}. Conductive opacities are those
of \citet{2007ApJ...661.1094C}. For the main sequence evolution, we
considered the equation of state from OPAL for H- and He-rich
compositions.  Also, we have taken from
\citet{1996ApJS..102..411I} the neutrino  emission   rates   for   pair,
 photo,   and bremsstrahlung processes, and for plasma processes we
included the treatment of   \citet{1994ApJ...425..222H}. For the WD
regime we have employed  an updated version of
the \citet{1979A&A....72..134M} equation of state. The nuclear network
takes into account 16 elements and 34 thermonuclear  reaction  rates
for  pp-chains,  CNO bi-cycle,  He burning, and C ignition. We also
considered time-dependent diffusion due to gravitational  settling and
chemical  and thermal diffusion of nuclear  species  following  the
multicomponent  gas  treatment  of
\citet{1969fecg.book.....B}. We have computed abundance changes
according to element diffusion, nuclear reactions,  and convective
mixing, a treatment  that represents a very significant aspect in
evaluating the importance of residual nuclear burning during the
cooling stage of low-mass WDs.

The pulsation analysis was carried out for radial ($\ell= 0$) and
non-radial ($\ell= 1, 2$) $p$ and $g$ modes, on the basis of
the set of adiabatic and non adiabatic pulsation periods presented
in \cite{2014A&A...569A.106C,2016A&A...585A...1C}, computed
employing the adiabatic and non adiabatic versions of the
{\tt LP-PUL} pulsation code
\citep{2006A&A...454..863C,2006A&A...458..259C}.
The adiabatic version of the {\tt LP-PUL} pulsation code is coupled
to the {\tt LPCODE} evolutionary code, and is  based  on  a  general
Newton-Raphson technique that  solves the fourth-order (second-order)
set  of real equations and  boundary conditions  governing   linear,
adiabatic, nonradial (radial) stellar  pulsations following the
dimensionless formulation
of \citet[][]{1971AcA....21..289D} \citep[see,
also,][]{1989nos..book.....U}. On the other side, the non-radial
(radial) non adiabatic version of the {\tt LP-PUL} pulsation code solves the
sixth-order (fourth-order) complex system of linearized equations and boundary
conditions as given by \cite{1989nos..book.....U} \citep{1983ApJ...265..982S}. Our
nonadiabatic computations rely on the frozen-convection
approximation, in which the perturbation of the convective flux
is neglected.
To compute the Brunt-V\"ais\"al\"a
frequency ($N$) we follow the so-called ``Ledoux Modified''
treatment \citep{1990ApJS...72..335T,1991ApJ...367..601B}.

Regarding the evolutionary sequences, realistic configurations for
low-mass He-core WD stars were derived by \cite{2013A&A...557A..19A}
by mimicking the binary evolution of progenitor stars. Binary
evolution was assumed to be fully nonconservative, and the losses of
angular momentum due to mass loss, gravitational wave radiation, and
magnetic braking were considered. All of the He-core WD initial models
were derived from evolutionary calculations for binary systems
consisting of an evolving Main Sequence low-mass component (donor
star) of initially $1 M_{\sun}$ and a $1.4 M_{\sun}$ neutron star
companion as the other component. A total of 14 initial He-core WD
models with stellar masses of  $0.1554$, $0.1612$, $0.1650$, $0.1706$,
$0.1762$, $0.1805$, $0.1863$, $0.1917$, $0.2019$,  $0.2389$,
$0.2707$,  $0.3205$, $0.3624$ and $0.4352\ M_{\sun}$ were computed for
initial orbital periods at the beginning of the Roche lobe phase in
the range of $0.9$ to $300\ $d.  The evolution of these models was
computed down to the range of luminosities of cool WDs, including the
stages of multiple thermonuclear CNO flashes during the beginning of
the cooling branch.

\section{Asteroseismological analysis}  
\label{analysis}  

Heretofore, asteroseismology has been applied to infer the fundamental
parameters of numerous pulsating WD stars.  Specifically, by
confronting the observed frequencies (or periods) of pulsating WDs
and appropriate theoretical models, it has been possible to infer
details about their origin, internal structure and evolution.  The
larger the number of frequencies detected in a given pulsating WD, the
more the information that can be inferred such as gravity, effective
temperature, stellar mass, and also  the internal chemical
stratification, the rate of rotation, the existence of magnetic
fields, the cooling timescale and core composition, among others. For
instance, the works of \cite{1998ApJS..116..307B},
\cite{2012MNRAS.420.1462R} and \cite{2016ApJS..223...10G,
2017A&A...598A.109G,2017ApJ...834..136G} have proven that
asteroseismology is a powerful technique to explore the interior of WDs.

In the next subsections, we describe the asteroseismological methods employed.

\subsection{Searching for a constant  period spacing}  
\label{estimation}

In the asymptotic limit of  high-radial order $k$, non-radial $g$ modes
with the same harmonic degree $\ell$ are expected to be equally spaced 
in period \citep{1980ApJS...43..469T}:

\begin{equation}
\label{eq:spacing}
\Delta\Pi^{\rm a}_{\ell}=\Pi_{k+1,\ell}-\Pi_{k,\ell}=\frac{2\pi^2}{\sqrt{\ell(\ell+1)}} \left[\int^{R_{\star}}_0{\frac{N(r)}{r}dr}\right]^{-1}
\end{equation}

where $N$  is   the  Brunt-V\"ais\"al\"a  frequency.  In principle,
the asymptotic period spacing or the average of the  computed period
spacings calculated from a grid of models (with different masses and
effective temperatures) can be compared with the mean period spacing
exhibited by a pulsating WD star, and then  a value of the stellar
mass can be inferred. The initial step to do so is to obtain (if it
exists) a mean period spacing underlying the observed periodicities.
We searched for  a constant  period  spacing  in  the data  of the
target stars     by  using  the Kolmogorov-Smirnov  \citep[K-S;
see][]{1988IAUS..123..329K},  the  Inverse Variance \citep[I-V;
see][]{1994MNRAS.270..222O} and the Fourier Transform \citep[F-T;
see][]{1997MNRAS.286..303H}  significance tests. In the  K-S  test,
the quantity $Q$ is  defined as the probability that  the observed
periods are   randomly  distributed.   Thus,  any   uniform  ---or
at  least systematically  non-random--- period  spacing present  in
the period spectrum of the star under analysis will appear as a
minimum in $Q$.  In the I-V test, a maximum  of the  inverse  variance
will indicate   the  presence of  a constant period spacing. Finally,
in the F-T test,  we calculate the Fourier Transform of a Dirac comb
function (created from a set of observed periods), and then we plot
the square of the amplitude of the resulting function in terms of
the inverse of the frequency. And once again, a maximum in the  square
of the amplitude will indicate the presence of a constant period
spacing.  

\begin{figure} 
\begin{center}
\includegraphics[clip,width=9 cm]{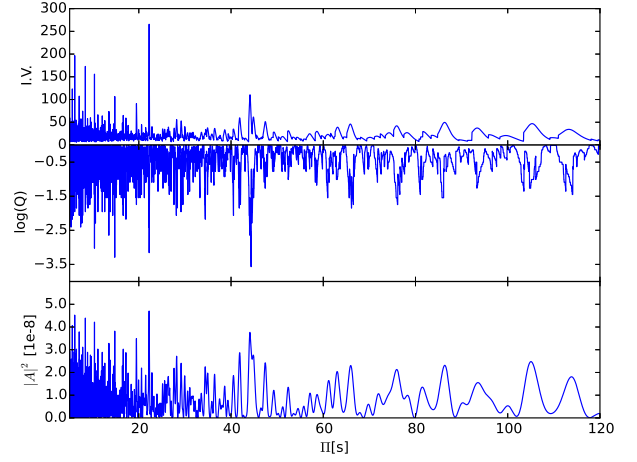} 
\caption{I-V (upper panel), K-S (middle panel), and F-T significance
(bottom panel) tests applied to the period spectrum of
J1518 to search for a constant period spacing. The periods
used here are those indicated in Table~\ref{table3}.}
\label{tests_j1518} 
\end{center}
\end{figure}

\begin{figure} 
\begin{center}
\includegraphics[clip,width=9 cm]{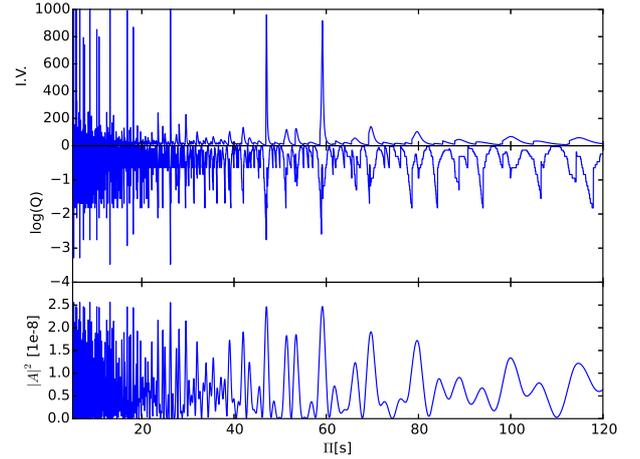} 
\caption{Same as Fig.~\ref{tests_j1518}, but for J1735. The periods
used here are those indicated in Table~\ref{table8}.} 
\label{tests_j1735} 
\end{center}
\end{figure}

As we can see in Tables~\ref{table1} to \ref{table9}, the number of
periods  exhibited by all the known ELMV WDs varies from 2 to 7. In
particular, because of the few periods exhibited by J1614, J2228,
J1738, J1618 and J2139, it is not possible to search for a constant
period spacing in these cases.  But in the cases of J1840, J1112,
J1518 and J1735 we were able to carry this procedure
out. Unfortunately, for the cases of J1840 and J1112, we could not
estimate any unambiguous constant period spacing, and furthermore,
there was not agreement between the significance tests.  It might be
due to the fact that the periods exhibited by these stars are not
fully in the asymptotic regime and/or there are not as many periods as
needed to determine a mean period spacing.  However, for the cases of
J1518 and J1735, as shown in Figs.~\ref{tests_j1518}
and \ref{tests_j1735}, respectively, we found clear indication of a
constant period spacing for the three independent significance tests
for both stars. For J1518, it lies at roughly  $\Delta \Pi \sim 44\
$s, though there is also another possible value at $\sim 22.2\ $s,
both for the three significance tests.  However, the latter is too
short and probably represents the harmonic of the main period spacing
($\frac{1}{2} \Delta \Pi$). In addition, a period separation of $\sim
22$ s is not likely to be the asymptotic period spacing because the
resulting stellar mass would be prohibitively high
\citep{2014A&A...569A.106C}. So, we assume that the period spacing associated
with J1518 is $\Delta \Pi^{\rm O} \sim 44\ $s. In the case of J1735, there is
a possible value for the period spacing at $\sim 26\ $s but once
again, as already mentioned, this value is too short and we discard
it. We can also see that there are two other possibilities at $\sim
47\ $s and $\sim 59\ $s, the latter being a more expectable value for
the period spacing, according to the asymptotic predictions. So, we
adopt $\Delta \Pi^{\rm O} \sim 59$ s as the period spacing for this star.

\subsection{Determination of the stellar mass
of J1518 and J1735 from the observed period spacing}  
\label{psp}  

In this section, we aim to estimate the masses of J1518 and J1735  by
comparing the average  of the
computed  period  spacings ($\overline{\Delta \Pi_{\ell}}$) for our grid
of models with  the observed period  spacing ($\Delta \Pi^{\rm O}_{\ell}$)
determined in the previous Section for each star. We warn that this
approach has an issue: the  period  spacing in this type of stars could
be sensitive also on the thickness of the outer H
envelope in addition to the stellar mass
\citep{1990ApJS...72..335T,2008PASP..120.1043F}. We defer to a future
publication a full exploration of this dependence. 

\begin{figure}  
\centering  
\includegraphics[clip,width=250pt]{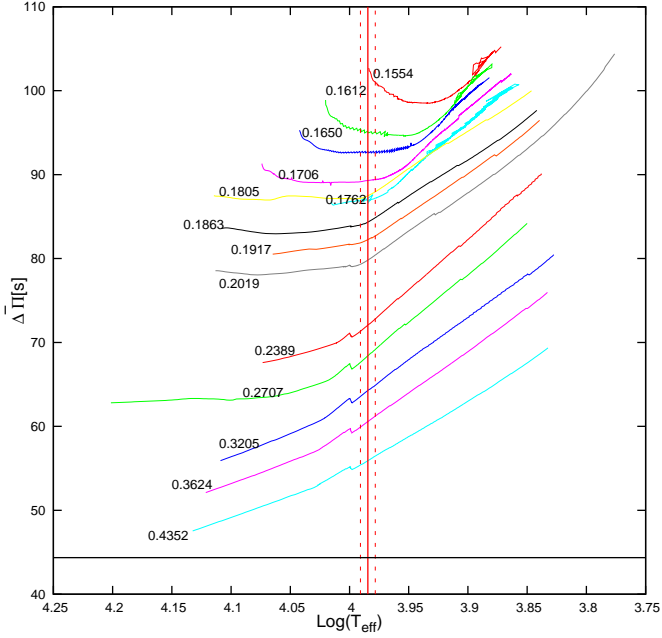}  
\caption{Average of the computed   
dipole ($\ell= 1$) period spacings ($\overline{\Delta \Pi_{\ell}}$) assessed 
in the range of the periods observed in J1518, corresponding
to each WD model sequence considered in this work, in terms of  the  logarithm 
of the effective temperature.  Numbers  along each  curve denote  
the  stellar mass (in solar  units). The observed period spacing
derived for J1518 is depicted with a horizontal solid line. Also indicated
are the $T_{\rm eff}$ (vertical solid line) in the 3D model, and its
uncertainties (vertical dashed lines).}
\label{prom_j1518}  
\end{figure}  

\begin{figure}  
\centering  
\includegraphics[clip,width=250pt]{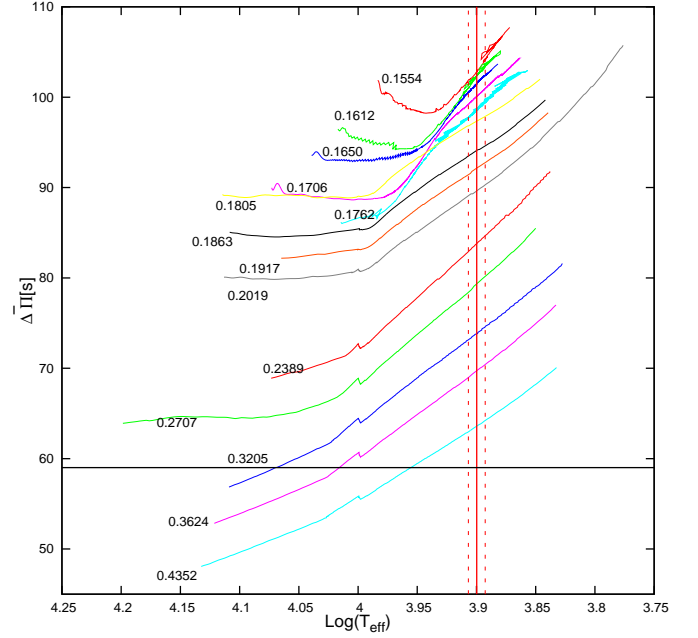}  
\caption{Same as Fig.~\ref{prom_j1518}, but for the case of J1735.}
\label{prom_j1735}  
\end{figure}  

The  average   of  the  computed   period  spacings  is   assessed  as
$\overline{\Delta  \Pi_{\ell}}= (n-1)^{-1}  \sum_{k}  \Delta \Pi_k  $,
where    the    ``forward''    period   spacing    is    defined    as
$\Delta \Pi_k= \Pi_{k+1}-\Pi_k$  ($k$ being the radial  order) and $n$
is the number  of theoretical periods within the range  of the periods
observed in  the target star.  For J1518, $\Pi_k \in  [1330,3900]\ $s,
while for J1735, $\Pi_k \in [3350,5000]\ $s.
  
In  Fig.~\ref{prom_j1518}  we show  the  run  of  the average  of  the
computed  period  spacings ($\ell=  1$)  for  J1518  in terms  of  the
logarithm of  the effective  temperature for  our ELM  WD evolutionary
sequences, along with the observed period spacing for J1518. As can be
seen    from    the    Figure,    the   smaller    the    values    of
$\overline{\Delta \Pi_{\ell}}$, the greater  the stellar mass. In this
case, it  clearly shows that  such a low value  ($\sim 44\ $s)  of the
observed  period spacing  would lead  to a  stellar mass  greater than
$0.4352\ M_{\sun}$,  which is  higher than expected  for this  type of
stars. Hence, this analysis does not seem to lead to a proper value of
the mass for J1518. It might be due  to the fact that this star is not
pulsating in the asymptotic regime
\citep[see][]{2014A&A...569A.106C}. 

In  Fig.~\ref{prom_j1735}  we show  the  run  of  the average  of  the
computed  period  spacings ($\ell=  1$)  for  J1735  in terms  of  the
logarithm of  the effective  temperature for  our ELM  WD evolutionary
sequences,  along with  the observed  period spacing  for J1735.  Once
again,  the value  we  obtain  for the  stellar  mass  is higher  than
expected  ($\gtrsim 0.43\  M_{\sun}$), even  though this  star may  be
pulsating in  the asymptotic  regime. Alternatively,  if the  value we
have obtained  for the observed period  spacing is real, in  the sense
that it can be associated with  the behaviour of high-radial order $g$
modes,  then it  would indicate  that this  star has  a mass  somewhat
larger  than   $0.4352\  M_{\sun}$,   and  that   their  spectroscopic
parameters ($T_{\rm eff}, \log g$) could be wrong.
\\

In the next Section we shall follow another approach to estimate the
stellar mass and other features of all the known ELMVs, through the
search of theoretical models that best fit the individual observed
periods. The advantage of this procedure is that, once we have chosen
a model, we have access to information of the star otherwise very
difficult (if not impossible) to know by any other method.

\subsection{Constraints from the individual observed periods: searching
for the best-fit model}  
\label{fitting}  

In this approach  we search for a pulsation model that best matches
the
\emph{individual} pulsation periods of a given star under study. The  
goodness of  the  match  between the   theoretical pulsation  periods
($\Pi_k^{\rm  T}$) and  the observed   individual  periods
($\Pi_i^{\rm  O}$) is measured by means  of a merit function defined
as: 

\begin{equation}
\label{chi}
\chi^2(M_{\star},  T_{\rm   eff})=   \frac{1}{m} \sum_{i=1}^{m}   \min[(\Pi_i^{\rm   O}-   \Pi_k^{\rm  T})^2], 
\end{equation}

\noindent where $m$ is the number of observed periods. The ELM model that
shows the lowest value of $\chi^2$, if exists, is adopted as the ``best-fit
model''.  We assess the function
$\chi^2=\chi^2(M_{\star}, T_{\rm eff})$ for stellar masses of $0.1554$,
$0.1612$, $0.1650$, $0.1706$, $0.1762$, $0.1805$, $0.1863$, $0.1917$,
 $0.2019$,  $0.2389$,  $0.2707$,  $0.3205$,  $0.3624$, and $0.4352 \ M_{\sun}$.
 For the effective temperature we also cover a wide range:
 $13000 \gtrsim T_{\rm eff} \gtrsim 6000\ $ K.

We have carried out asteroseismological fits for all the known ELMV WD
stars.  This is the first time that this procedure is used for this
type of stars. We start our analysis assuming that  all of  the
observed periods correspond to $g$ modes associated with $\ell= 1$,
and considering the set of observed periods,  $\Pi_i^{\rm  O}$, of
each star to compute the quality function given by
Eq.~(\ref{chi}). Next, we considered the case in which  all of  the
observed periods correspond to $g$ modes associated with $\ell=  2$,
and finally, we considered the case of a mix of $g$ modes
associated with $\ell= 1$ and $\ell= 2$. In the case of J1112, we
performed a more detailed analysis. For this star we worked with two
different sets of observed periods. On the one hand, the five longest
periods, for which we carried out the analysis previously
mentioned. On the other hand, we adopted the whole set of periods
(seven) and explored the possibility that they correspond to a mix
of $g$ and $p$ modes with $\ell = 1$, and also we considered the case
in which the observed periods correspond to radial ($\ell= 0$) and $p$
and $g$ modes ($\ell = 1, 2$).

Figs.~\ref{fig:ajustej1840} to~\ref{fig:ajustej2139} show the quantity
$(\chi^2)^{-1}$ in terms of the effective temperature  for different
stellar  masses, for each known ELMV, taking into account the
corresponding set of observed periods. We also
include the effective temperatures and their uncertainties for the
1D (dark gray vertical lines) and 3D model atmospheres (red vertical lines)
determinations. As mentioned before, the goodness of the match between the 
theoretical and the observed periods is measured by the value of
$\chi^2$: the better the period match, the lower the value of $\chi^2$ 
---in the figures, the greater the value of $(\chi^2)^{-1}$. In some cases,
there is a single maximum and we adopt that model as the asteroseismological
solution for that star. Sometimes, there are multiple possible solutions,
and we need to employ some external constraint in order to choose one.
Generally, the constraint is the uncertainty in the effective temperature,
given by the spectroscopy. In some cases, when there are still multiple
possible solutions, we choose the model with a mass as close as possible to the
mass given by the spectroscopic determinations. It is important to
mention that, as it is more likely to observe $\ell= 1$
than $\ell= 2$ modes \citep[due to geometric
cancellation effects which become stronger with higher values of $\ell$; see][]
{1977AcA....27..203D}, we will usually choose, \emph{when possible},
the asteroseismological solutions that fit to observed periods with a larger
number of $\ell= 1$ modes. 

\subsubsection{The case of J1840}

\begin{figure}  
\centering  
\includegraphics[clip,width=250pt]{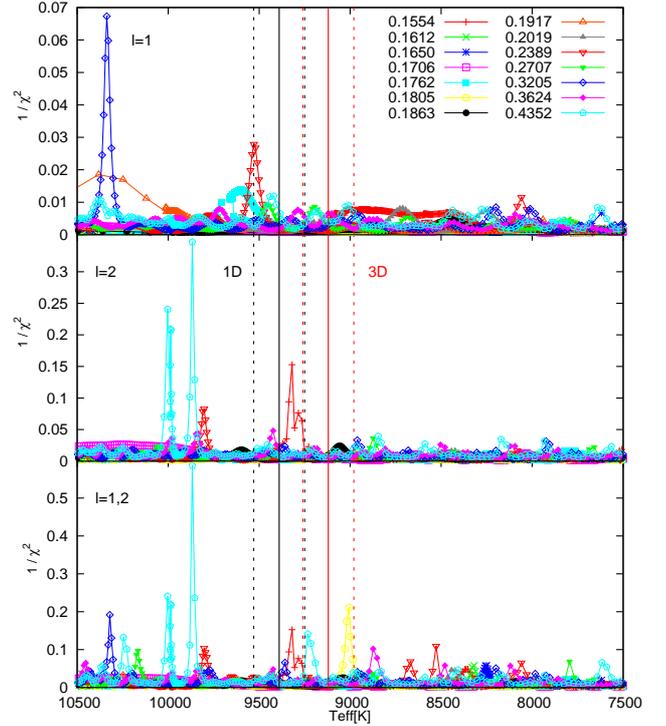}  
\caption{Inverse of the quality function of the period fit considering
    $\ell= 1$ (top panel), $\ell= 2$ (middle panel) and $\ell= 1, 2$ (bottom
    panel) versus $T_{\rm  eff}$, for J1840 (see text for details).
    The vertical strips depict the spectroscopic $T_{\rm  eff}$ (solid vertical
    lines)
    and their uncertainties (dashed vertical lines) for the 1D (dark gray lines)
    and 3D models (red lines)
    [Color figure only available in the electronic version of the article].}
\label{fig:ajustej1840}
\end{figure}  

In Fig.~\ref{fig:ajustej1840},  the match between the  theoretical and
the five observed periods of J1840, assuming they are associated with
$g$ modes, for the cases of $\ell= 1$ (top panel), $\ell= 2$ (middle
panel) and $\ell= 1, 2$ (bottom panel), is shown. The case $\ell= 2$
is only depicted for the seek of completeness, since we do not expect
that a pulsating
star can exhibit all the periods associated to $\ell= 2$ and none of
them corresponding to $\ell= 1$, due to geometric cancellation arguments
(see above).  For the $\ell= 1$
case, the upper panel shows that there is more than one solution. In
particular, the absolute maximum (the best solution) lies at a very
higher effective temperature than that allowed by both spectroscopic
determinations. The second best solution lies within the range of
allowed $T_{\rm eff}$ for the 1D model atmosphere ($T_{\rm eff}=
9390 \pm 140\ $K), so we may adopt this model because it represents a
good period fit. It corresponds to a mass of
$M_{\star}= 0.2389\ M_{\sun}$ at $T_{\rm  eff} \sim 9529$ K.
For the $\ell= 1, 2$ case,
once again there is no a single solution, and the best period fit has
a very high value of $T_{\rm  eff}$.  In the ranges of allowed $T_{\rm
eff}$, we can see that there is no unambiguous asteroseismological
model. However, we may adopt the solution given for the model with
$M_{\star}= 0.1805\ M_{\sun}$ at $T_{\rm  eff} \sim 9007$ K, which is
the best fit in the range of allowed $T_{\rm  eff}$ for the 3D model
atmosphere ($T_{\rm eff}= 9120 \pm 140\ $K).

In order to know how good the agreement of theoretical and observed
periods is, we can compare them by computing the absolute period
differences $|\delta\Pi|= |\Pi^{\rm O}-\Pi^{\rm T}|$.  The results for
J1840 are shown in Table~\ref{tab:perj1840_a}, for the case of $\ell=
1$. Column 6 of Table \ref{tab:perj1840_a} shows the value of the
linear nonadiabatic growth rate ($\eta$), defined as $\eta$ ($\equiv
-\Im(\sigma)/ \Re(\sigma)$,  where $\Re(\sigma)$  and $\Im(\sigma)$
are the real and the  imaginary part, respectively,  of the complex
eigenfrequency  $\sigma$, computed  with the nonadiabatic version of
the {\tt LP-PUL} pulsation code 
\citep{2006A&A...458..259C,2016A&A...585A...1C}. A value of $\eta > 0$ ($\eta < 0$) 
 implies an unstable (stable) mode (see column 6 of
Table \ref{tab:perj1840_a}). For the case of $\ell= 1,2$ the results
are shown in Table~\ref{tab:perj1840_b}.

Considering that the period fit of the $\ell= 1, 2$ case is better,
since it has a larger value of $(\chi^2)^{-1}$ than the solution of
the $\ell= 1$ case, and the mass of this model ($M_{\star}= 0.1805\
M_{\sun}$) is in line with the spectroscopic masses determined for
this star ($M^{(1D)}_{\star}= 0.183\ M_{\sun}$ and $M^{(3D)}_{\star}=
0.177\ M_{\sun}$), we adopt the model with  $M_{\star}= 0.1805\
M_{\sun}$ and $T_{\rm eff}= 9007\ $K as the asteroseismological
solution for J1840, which has a value of $T_{\rm eff}$ in agreement
with the spectroscopy (for the 3D model), even though this model
has a larger number of $\ell= 2$ than $\ell= 1$ modes. Note that
most of the periods of this model correspond to pulsationally unstable modes.

\begin{table}[ht]
\centering
\caption{Comparison of the observed and theoretical periods for
J1840, corresponding
to the asteroseismological model with $M_{\star}= 0.2389\ M_{\sun}$
and $T_{\rm eff}= 9529\ $ K ($\ell= 1$).  Also shown are the harmonic
degree $\ell$, the radial order $k$, the absolute period difference,
and the nonadiabatic growth rate for each theoretical period.}
\begin{tabular}{ccccccc}
\hline
\hline
 $\Pi^{\rm O}$[s] & $\Pi^{\rm T}$[s] & $\ell$ & $k$ & $|\delta\Pi|$[s] & $\eta[10^{-5}]$ & 
Remark\\
\hline
$1164.15$& $1168.26$ & 1 & $14$ &  $4.11$  & $0.0345$ & unstable \\
$1578.70$& $1589.47$ & 1 & $20$ &  $10.77$ & $0.384$ & unstable \\
$2376.07$& $2378.49$ & 1 & $31$ &  $2.42$  & $2.61$ & unstable \\
$3930.0$ & $3923.65$ & 1 & $52$ &  $6.35$  & $4.72$ & unstable \\
$4445.3$ & $4445.20$ & 1 & $59$ &  $0.1$   & $4.16$ & unstable \\
\hline
\hline
\end{tabular}
\label{tab:perj1840_a}
\end{table}

\begin{table}[ht]
\centering
\caption{Same as Table~\ref{tab:perj1840_a}, but for the model
with $M_{\star}= 0.1805\ M_{\sun}$ and $T_{\rm eff}= 9007\ $K,
in the case of $\ell= 1,2$, adopted for J1840.} 
\begin{tabular}{cccccccc}
\hline
\hline
 $\Pi^{\rm O}$[s] & $\Pi^{\rm T}$[s] & $\ell$ & $k$ & $|\delta\Pi|$[s] & $\eta[10^{-6}]$ &
Remark\\
\noalign{\smallskip}
\hline
$1164.15$& $1163.53$& $2$ & $20$&  $0.62$ & $0.719$ & unstable \\
$1578.70$& $1577.75$& $2$ & $28$&  $0.95$ & $2.47$ & unstable \\
$2376.07$& $2373.77$& $2$ & $43$&  $2.30$ & $4.13$ & unstable \\ 
$3930.0$& $3933.98$& $2$ & $72$&   $3.98$ & $-3.22$& stable \\
$4445.3$& $4444.16$& $1$ & $47$&   $1.14$ & $6.96$  & unstable \\
\hline
\hline
\end{tabular}
\label{tab:perj1840_b}
\end{table}

\subsubsection{The case of J1112}

\begin{figure}  
\centering  
\includegraphics[clip,width=250pt]{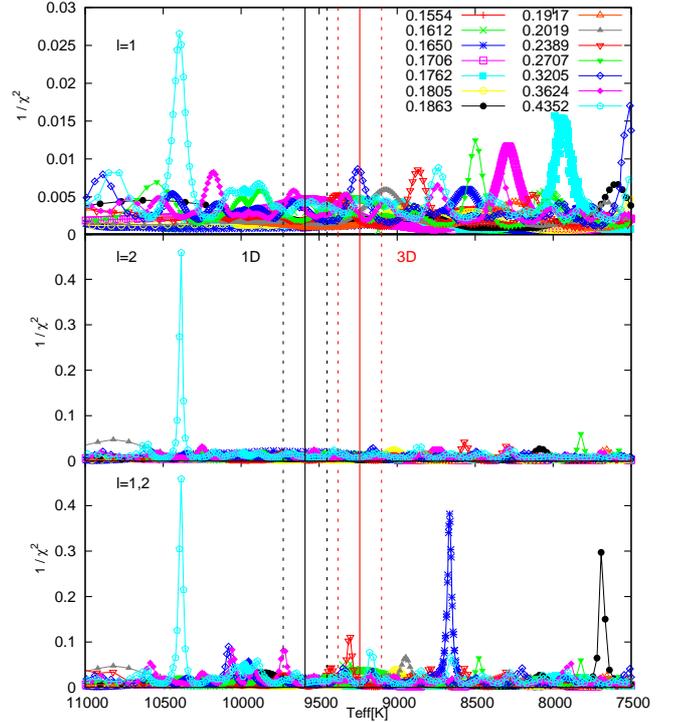}  
\caption{Same as Fig.~\ref{fig:ajustej1840}, but for the case of J1112 for the
five longest periods.}
\label{fig:ajustej1112g}
\end{figure}  

Due to the presence of the two short periods at $\sim 108$ s and $\sim
134$ s in the pulsation spectrum of J1112, which are probably
associated with $p$ or radial modes, we divided the analysis for this
star into two parts: on the one hand, we considered the five longest
observed periods, assuming that they are all associated with $\ell=
1$, $\ell= 2$ or a mix of $\ell= 1$ and $\ell= 2$ $g$ modes.On
the other hand, we considered the whole set of periods (seven),
considering firstly, that they are associated with a mix of $\ell=
1$ $p$ and $g$ modes\footnote{We have
also explored a possible combination of $g$ and $p$ modes with $\ell= 1$ and
$\ell= 2$, but we have not obtained significantly different results.},
and secondly, with a mix of $p$ and $g$
modes with $\ell= 1$ and $\ell= 2$, and also radial modes ($\ell=0$).
At this point, it is worth mentioning that the reality of
these two short periods are not definitively confirmed, as claimed
by \cite{2013ApJ...765..102H}. Thus, these separated analyses are worth
doing.

In Fig.~\ref{fig:ajustej1112g} we show the match between the 
theoretical and the five longest observed periods of J1112, assuming they
are associated with $g$ modes, for the cases of $\ell= 1$ (top panel),
$\ell= 2$ (middle panel, shown for completeness) and $\ell= 1, 2$
(bottom panel). In the case of $\ell= 1$, we
can see that the absolute maximum is at a very high effective temperature, and
there are also many other solutions for low values of $T_{\rm  eff}$.
Within the range of allowed $T_{\rm  eff}$ for the 3D model atmosphere analysis
($T_{\rm eff}= 9240 \pm 140\ $K), there may be a
solution for the model with $M_{\star}= 0.3205\ M_{\sun}$ at
$T_{\rm  eff} \sim 9253$ K. We compare the observed and the theoretical
periods as we did for the previous star, and the results are displayed in
Table~\ref{tab:perj1112_a}. If we consider the case of $\ell= 1, 2$,
there are multiple local maxima which are either too hot or too cold
in comparison with the allowed values
of $T_{\rm  eff}$. Nevertheless, there may be a possible solution within
the range of allowed $T_{\rm  eff}$ (for the 3D model), for the case of
$M_{\star}= 0.2389\ M_{\sun}$ at $T_{\rm  eff} \sim 9300$ K. In
Table~\ref{tab:perj1112_b} we show the comparison between the
observed and the theoretical periods for this model.

\begin{table}[ht]
\centering
\caption{Same as Table~\ref{tab:perj1840_a}, but for J1112
(in the case of the five longest periods) for the model with
$M_{\star}= 0.3205\ M_{\sun}$ and
$T_{\rm eff}= 9253$ K, in the case of $\ell= 1$. }
\begin{tabular}{ccccccc}
\hline
\hline
 $\Pi^{\rm O}$[s] & $\Pi^{\rm T}$[s] & $\ell$ & $k$ &
 $|\delta\Pi|$[s] & $\eta[10^{-6}]$ & 
Remark\\
\hline
$1792.905$& $1802.269$& 1 & $26$&  $9.364$ & $4.90$ & unstable \\
$1884.599$& $1867.419$& 1 & $27$&  $17.18$ & $4.87$ & unstable \\
$2258.528$& $2264.984$& 1 & $33$&  $6.456$ & $5.54$ & unstable \\
$2539.695$& $2530.317$& 1 & $37$&  $9.378$ & $5.91$ & unstable \\
$2855.728$& $2863.702$& 1 & $42$&  $7.974$ & $5.08$ & unstable \\
\hline
\hline
\end{tabular}
\label{tab:perj1112_a}
\end{table}

\begin{table}[ht]
\centering
\caption{Same as Table~\ref{tab:perj1112_a} for J1112 (in the case of the five
longest periods), but for the model with $M_{\star}= 0.2389\ M_{\sun}$ and
$T_{\rm eff}= 9300$ K, in the case of $\ell= 1, 2$.} 
\begin{tabular}{cccccccc}
\hline
\hline
 $\Pi^{\rm O}$[s] & $\Pi^{\rm T}$[s] &  $\ell$ & $k$ & $|\delta\Pi|$[s] & $\eta[10^{-6}]$ &
Remark\\
\noalign{\smallskip}
\hline
$1792.905$& $1798.677$& $2$ & $40$&  $5.772$ & $6.71$ & unstable \\
$1884.599$& $1884.824$& $2$ & $42$&  $0.225$ & $6.63$ & unstable \\
$2258.528$& $2259.902$& $1$ & $29$&  $1.374$ & $7.13$ & unstable \\
$2539.695$& $2536.648$& $2$ & $57$&  $3.047$ & $3.79$ & unstable \\
$2855.728$& $2856.498$& $1$ & $37$&  $0.77$  & $11.2$ & unstable \\
\hline
\hline
\end{tabular}
\label{tab:perj1112_b}
\end{table}

\begin{figure}  
\centering  
\includegraphics[clip,width=250pt]{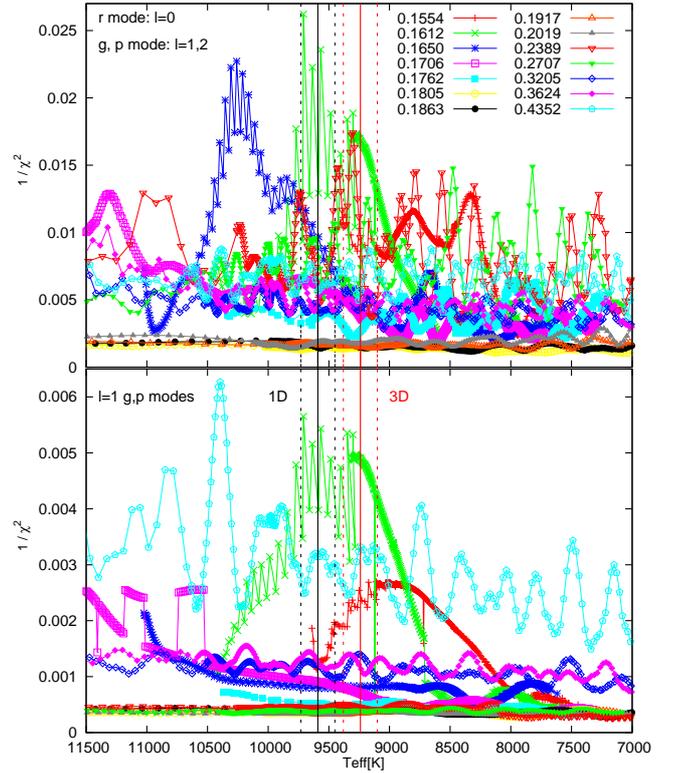}  
\caption{Same as Fig.~\ref{fig:ajustej1112g}, but for the case of the seven
observed periods of J1112. In the upper panel, the case of radial
($\ell= 0$) and $p$ and $g$ modes ($\ell= 1,2$) is shown, while in
the bottom panel the case of $p$ and $g$ modes with $\ell= 1$ is displayed.}
\label{fig:ajustej1112_gpr}
\end{figure}

Next, we considered the case in which the whole set of
observed periods (seven) corresponds to $p$ and $g$ modes with $\ell= 1$, and
also, the case in which it corresponds to radial ($\ell= 0$) and $p$ and $g$
modes ($\ell= 1,2$). The
results are shown in Fig. \ref{fig:ajustej1112_gpr}. For the case of the mix
of $p$ and $g$ modes with $\ell$= 1 (bottom panel), the absolute maximum lies
at a high value of $T_{\rm eff}$ and there is no unambiguous solution in the
allowed ranges of $T_{\rm eff}$. However, there may be one possible solution for
$M_{\star}= 0.1612\ M_{\sun}$, that lies in the allowed range of $T_{\rm eff}$ for the 1D
model ($T_{\rm eff}= 9590 \pm 140\ $K). In a more complete analysis, considering the
mix of  radial ($\ell= 0$) and $p$ and $g$ modes ($\ell= 1, 2$), we find
that the absolute maximum is at the same model and it represents a better
match because it has a larger value of $(\chi^2)^{-1}$, as shown in the top panel
of Fig.\ref{fig:ajustej1112_gpr}. It is the best period fit for this case, and
corresponds to
$M_{\star}= 0.1612\ M_{\sun}$ at $T_{\rm  eff} \sim 9709$ K. Hence, this represents a
very good solution for the case of the whole set of periods. Once
again, we show the comparison between the observed and the theoretical periods
in Table~\ref{tab:perj1112_c}. We can see from this Table that one of the short
periods may be associated with a $p$ mode (with $\ell= 2$) and the other one, with
a radial mode ($\ell= 0$).

Considering all the results from this analysis, we may conclude that the
best solution corresponds to the model with $M_{\star}= 0.1612\ M_{\sun}$ and
$T_{\rm eff}= 9709$ K, which is quite in line with the spectroscopic masses determined for
this star ($M^{(1D)}_{\star}= 0.179\ M_{\sun}$ and $M^{(3D)}_{\star}=
0.169\ M_{\sun}$) and  also, in line with the $T_{\rm eff}$ given by
the spectroscopy (for the 1D model). Unfortunately, our nonadiabatic
computations (see Table~\ref{tab:perj1112_c}) predict that all the modes of this
possible solution are pulsationally stable, forcing us to discard this solution.
If we consider this and ignore the two shortest periods of this star, we could
adopt the solution found
with a mass of $M_{\star}= 0.2389\ M_{\sun}$ and $T_{\rm eff}= 9300$ K
(see Table~ \ref{tab:perj1112_b}), although this value of the
stellar mass is not in such good agreement with the masses resulting from the
spectroscopy. Note, however, that all the periods of this model are associated with
pulsationally unstable modes. Finally, the fact that we are not able to find an
asteroseismological model having unstable modes with periods that fit
the seven periods observed in J1112 (including the shortest ones)
could be indicating that the periods at $\sim 108$ s and $\sim 134$ s
reported by \cite{2013ApJ...765..102H} could not be real. This calls for the need of
further photometric work on this star.

\begin{table*}[t]
\centering
\caption{Same as Table~\ref{tab:perj1112_a}, for J1112 (considering the
whole set of periods) but for the model with $M_{\star}= 0.1612\ M_{\sun}$
 and $T_{\rm eff}= 9709\ $K, in the case of $p$, $g$ ($\ell= 1,2$) and
 radial modes ($\ell$= 0).} 
\begin{tabular}{ccccccccc}
\hline
\hline
 $\Pi^{\rm O}$[s] & & $\Pi^{\rm T}$[s] &  &  $\ell$ & $k$ & $|\delta\Pi|$[s] & $\eta[10^{-7}]$ & 
Remark\\
\hline
\noalign{\smallskip}
 &                  $g$ & $p$ & radial        \\
\noalign{\smallskip}
\hline
$107.56$&    --- &    --- &  $105.176$ & $0$ & $1$&  $2.384$ & $-0.287$ & stable \\
$134.275$&    --- &    $136.881$ & --- & $2$ & $0$&  $2.606$ & $-0.0238$ & stable \\
$1792.905$&  $1793.216$&  --- & --- & $1$ & $17$&  $0.311$ &  $-0.0197$ & stable \\
$1884.599$& $1889.869$&  --- & --- & $2$ & $32$&  $5.270$  &  $-7.34$ & stable \\
$2258.528$& $2272.008$&  --- & --- & $2$ & $39$&  $13.480$ &  $-53.6$ & stable \\
$2539.695$& $2543.853$&  --- & --- & $2$ & $44$&  $4.158$ &   $-196$ & stable \\
$2855.728$& $2850.465$&  --- & --- & $1$ & $28$&  $5.263$ &   $-2.52$ & stable \\
\hline
\hline
\end{tabular}
\label{tab:perj1112_c}
\end{table*}

\subsubsection{The case of J1518}

\begin{figure}  
\centering  
\includegraphics[clip,width=250pt]{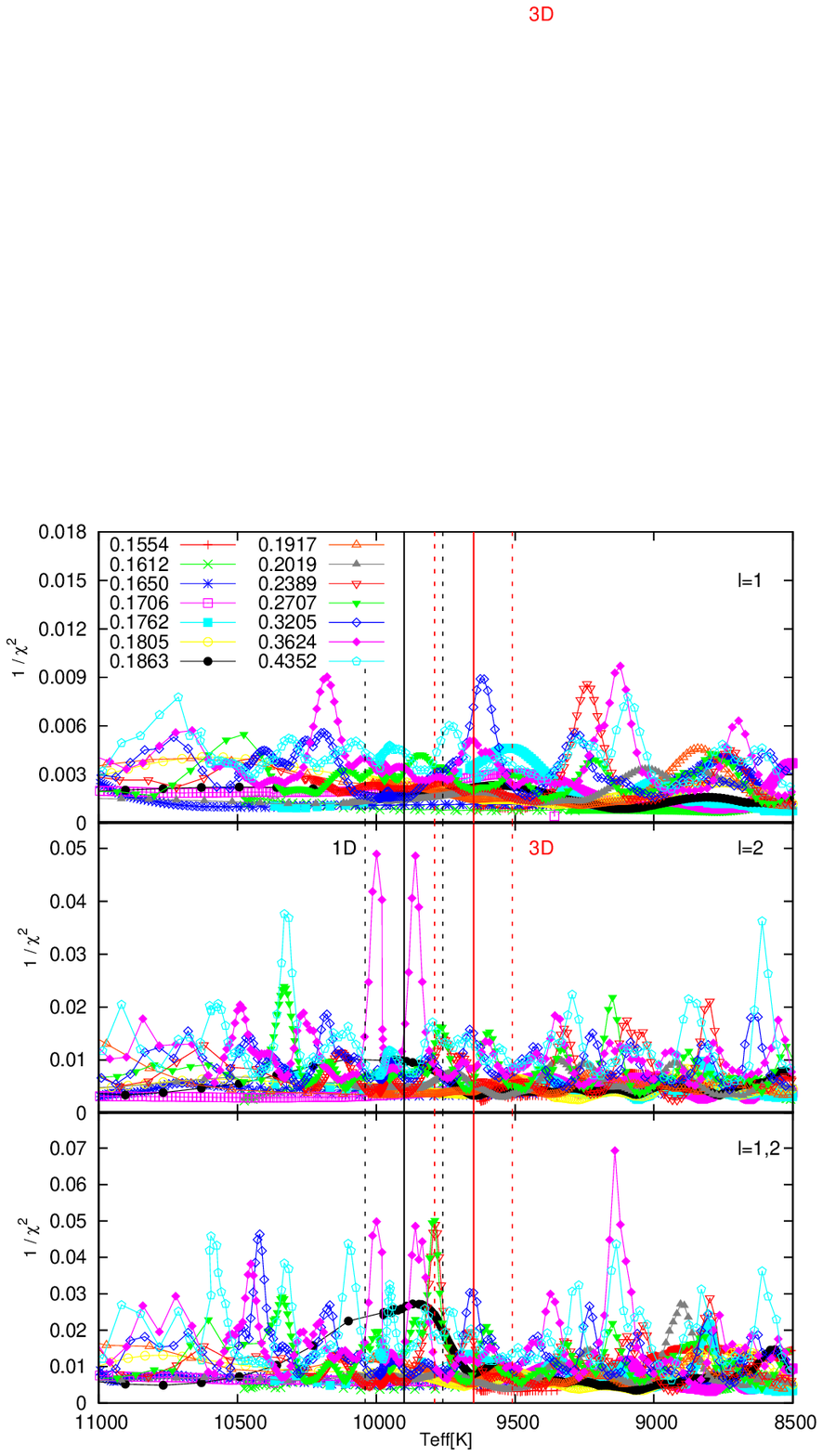}  
\caption{Same as Fig.~\ref{fig:ajustej1840}, but for the case of J1518.}
\label{fig:ajustej1518}
\end{figure}  

In Fig.~\ref{fig:ajustej1518} we show the match between the
theoretical periods and the seven observed periods of J1518 assuming
they are associated with $g$ modes, for the cases of $\ell= 1$ (top
panel), $\ell= 2$ (middle panel) and $\ell= 1, 2$ (bottom panel). In
the $\ell= 1$ case, we see multiple local maxima. However, in the
range of allowed $T_{\rm  eff}$ given by the 3D model atmosphere
calculations ($T_{\rm eff}= 9650 \pm 140\ $K), there is a possible
solution that may be chosen as a representative model for J1518. This
corresponds to $M_{\star}= 0.3205\ M_{\sun}$ at
$T_{\rm  eff} \sim 9625\ $K. In Table~\ref{tab:perj1518_a} we compare the
theoretical and the observed periods for this model.  For the case of
$\ell= 1,2$, the best period fit lies at a very low $T_{\rm  eff}$,
and there are multiple local maxima within the ranges of allowed
$T_{\rm  eff}$. However, there is a possible solution characterized by
$M_{\star}= 0.2707\ M_{\sun}$ and $T_{\rm  eff} \sim 9789$ K which,
although not so clear in the Figure, is the best period fit that lies
within the ranges of allowed $T_{\rm eff}$ ($T^{(1D)}_{\rm eff}=
9900 \pm 140\ $K and $T^{(3D)}_{\rm eff}= 9650 \pm 140\ $K). We
present in Table~\ref{tab:perj1518_b} the comparison between the
observed and the theoretical periods for this case.

Considering that the solution for the $\ell= 1, 2$ case implies a much
better fit (a larger value of $(\chi^2)^{-1}$) than the one for the
$\ell= 1$ case, we may adopt the model with $M_{\star}= 0.2707\
M_{\sun}$ and $T_{\rm eff}= 9789\ $K as the asteroseismological
solution for this star, which is in line with the $T_{\rm eff}$ given
by the spectroscopy. Moreover, most of the periods of the adopted model are
associated with pulsationally unstable modes. It is necessary to stress,
however, that none of the solutions found are in good agreement with the
masses resulting from the spectroscopic determinations
($M^{(1D)}_{\star}= 0.220\ M_{\sun}$ and $M^{(3D)}_{\star}= 0.197\ M_{\sun}$).

\begin{table}[ht]
\centering
\caption{Same as Table~\ref{tab:perj1840_a}, but for J1518 for the model
with $M_{\star}= 0.3205\ M_{\sun}$ and 
$T_{\rm eff}= 9625\ $K, in the case of $\ell= 1$. }
\begin{tabular}{ccccccc}
\hline
\hline
 $\Pi^{\rm O}$[s] & $\Pi^{\rm T}$[s] & $\ell$ & $k$ & $|\delta\Pi|$[s] & $\eta[10^{-5}]$ & 
Remark\\
\hline
$1335.318$& $1324.926$& 1 & $19$&  $10.392$ & $0.484$ & unstable \\
$1956.361$& $1953.996$& 1 & $29$&  $2.365$ & $1.71$ & unstable \\
$2134.027$& $2146.419$& 1 & $32$&  $12.392$ & $2.26$ & unstable \\
$2268.203$& $2275.543$& 1 & $34$&  $7.340$ & $2.46 $ & unstable \\
$2714.306$& $2727.475$& 1 & $41$&  $13.169$ & $2.49$ & unstable \\
$2799.087$& $2791.464$& 1 & $42$&  $7.623$ & $2.57 $ & unstable \\
$3848.201$& $3832.927$& 1 & $58$&  $15.274$ & $0.0899$ & unstable \\
\hline
\hline
\end{tabular}
\label{tab:perj1518_a}
\end{table}

\begin{table}[ht]
\centering
\caption{Same as Table~\ref{tab:perj1518_a}, but for the model adopted for J1518
with $M_{\star}= 0.2707\ M_{\sun}$ and 
$T_{\rm eff}= 9789$ K, in the case of $\ell= 1, 2$. }
\begin{tabular}{cccccccc}
\hline
\hline
 $\Pi^{\rm O}$[s] & $\Pi^{\rm T}$[s] &  $\ell$ & $k$ & $|\delta\Pi|$[s] & $\eta[10^{-5}]$ &
Remark\\
\noalign{\smallskip}
\hline
$1335.318$& $1331.485$& $2$ & $32$&  $3.833$ & $3.94$ & unstable \\
$1956.361$& $1960.394$& $2$ & $48$&  $4.033$ & $5.37$ & unstable \\
$2134.027$& $2140.805$& $1$ & $30$&  $6.778$ & $5.25$ & unstable \\
$2268.203$& $2274.699$& $1$ & $32$&  $6.496$ & $6.55$ & unstable \\
$2714.306$& $2714.827$& $2$ & $67$&  $0.521$ & $0.711$ & unstable \\
$2799.087$& $2794.753$& $2$ & $69$&  $4.334$ & $-0.357$ & stable \\
$3848.201$& $3847.023$& $1$ & $55$&  $1.178$ & $7.81$ & unstable \\
\hline
\hline
\end{tabular}
\label{tab:perj1518_b}
\end{table}

\subsubsection{The case of J1614}

\begin{figure}  
\centering  
\includegraphics[clip,width=250pt]{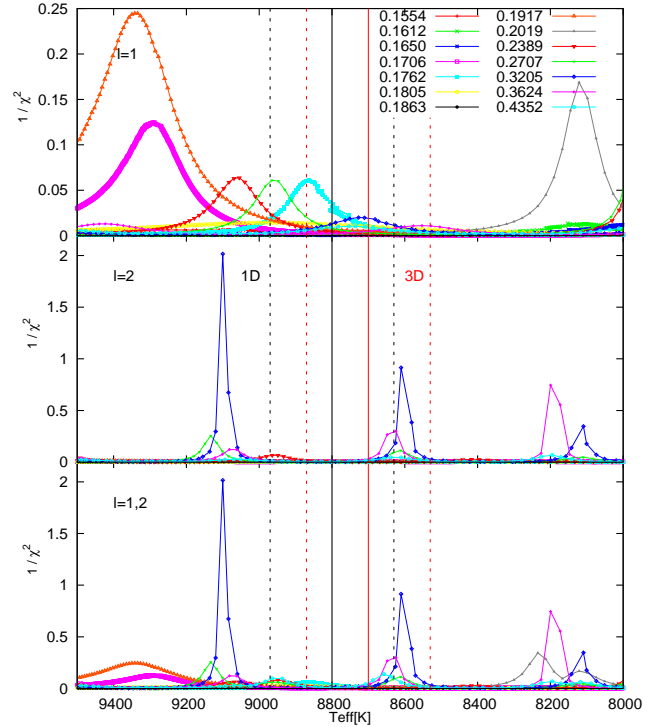}  
\caption{Same as Fig.~\ref{fig:ajustej1840}, but for the case of J1614.}
\label{fig:ajustej1614}
\end{figure}

In Fig.~\ref{fig:ajustej1614} we depict the match between the 
theoretical and the two observed periods of J1614, assuming they
are associated with $g$ modes, for the cases of $\ell= 1$ (top panel),
$\ell= 2$ (middle panel) and $\ell= 1, 2$ (bottom panel). It is worth
mentioning in advance that this period fit will not be robust because this star
only shows two independent periods.

In the case of $\ell= 1$, we can see that there is no unambiguous solution, and
the best solutions are located beyond the ranges of allowed $T_{\rm eff}$.
However, we may choose the model with
$M_{\star}= 0.1762\ M_{\sun}$ and $T_{\rm eff} \sim 8862\ $K that lies within the
ranges of allowed $T_{\rm eff}$ ($T^{(1D)}_{\rm eff}= 8800 \pm 170\ $K and
$T^{(3D)}_{\rm eff}= 8700 \pm 170\ $K), and also has a mass value consistent with the
spectroscopic determinations. In Table~\ref{tab:perj1614_a} we show the
comparison between the observed and the theoretical periods for this model.
In the case of $\ell= 1, 2$, the best fit is located at a high value of
$T_{\rm  eff}$, but the second best fit lies within the range of allowed
$T_{\rm  eff}$ (for the 3D model). It is characterized by $M_{\star}= 0.3205\ M_{\sun}$,
at $T_{\rm  eff} \sim 8610$ K. However, as can be seen in
Table~\ref{tab:perj1614_b}, the comparison between the observed and the
theoretical periods shows that both periods are associated with
$\ell= 2$, which is not usually the case because, as already stated,
it is more likely to observe periods associated with $\ell= 1$ than $\ell= 2$.
Hence, the asteroseismological model we adopt corresponds to the case of $\ell= 1$,
with $M_{\star}= 0.1762\ M_{\sun}$ and $T_{\rm eff}= 8862$ K, with a mass
in line with the spectroscopic determinations
($M^{(1D)}_{\star}= 0.192\ M_{\sun}$ and $M^{(3D)}_{\star}= 0.172\ M_{\sun}$) and a
$T_{\rm  eff}$ in agreement with the spectroscopy. Finally, as can be seen from
Table~\ref{tab:perj1614_a}, both periods are associated with pulsationally
unstable modes.

\begin{table}[ht]
\centering
\caption{Same as Table~\ref{tab:perj1840_a}, but for the model adopted for J1614 with
$M_{\star}= 0.1762\ M_{\sun}$ and $T_{\rm eff}= 8862\ $K, in the case of
$\ell= 1$.}
\begin{tabular}{cccccccc}
\hline
\hline
 $\Pi^{\rm O}$[s] & $\Pi^{\rm T}$[s] & $\ell$ & $k$ & $|\delta\Pi|$[s] & $\eta[10^{-9}]$ & 
Remark\\
\hline
$1184.106$& $1179.527$& 1 & $11$&  $4.579$ & $2.03$ & unstable \\
$1262.668$& $1266.119$& 1 & $12$&  $3.451$ & $2.88$ & unstable \\
\hline
\hline
\end{tabular}
\label{tab:perj1614_a}
\end{table}

\begin{table}[t]
\centering
\caption{Same as Table~\ref{tab:perj1614_a} for J1614, but for the model
with $M_{\star}= 0.3205\ M_{\sun}$ and $T_{\rm eff}= 8610\ $K, in the
case of $\ell= 1, 2$. }
\begin{tabular}{cccccccc}
\hline
\hline
 $\Pi^{\rm O}$[s] & $\Pi^{\rm T}$[s] &  $\ell$ & $k$ & $|\delta\Pi|$[s] & $\eta[10^{-7}]$ & 
Remark\\
\noalign{\smallskip}
\hline
$1184.106$& $1182.674$& $2$& $28$&  $1.432$ & $4.41$ & unstable \\
$1262.668$& $1263.035$& $2$& $30$&  $0.367$ & $4.13$ & unstable \\
\hline
\hline
\end{tabular}
\label{tab:perj1614_b}
\end{table}

\subsubsection{The case of J2228}

\begin{figure}  
\centering  
\includegraphics[clip,width=250pt]{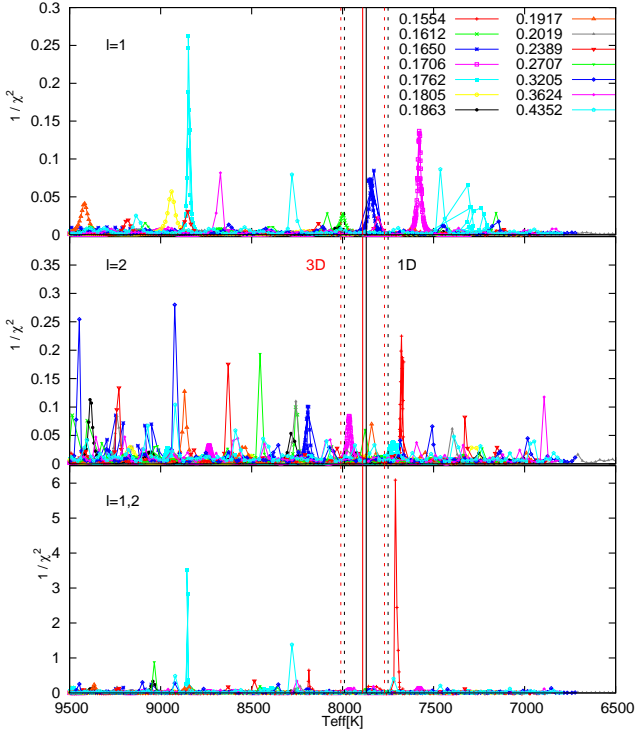}  
\caption{Same as Fig.~\ref{fig:ajustej1840}, but for the case of J2228.}
\label{fig:ajustej2228}
\end{figure}  

In Fig.~\ref{fig:ajustej2228} we can see the match between the
theoretical and the three observed periods of J2228, assuming they are
associated with $g$ modes, for the cases of $\ell= 1$ (top panel),
$\ell= 2$ (middle panel) and $\ell= 1, 2$ (bottom panel). In the case
of $\ell= 1$, we can see that there are multiple possible solutions
and that the best fit solution is located at a high value of
$T_{\rm eff}$. Within the ranges of allowed $T_{\rm  eff}$ ($T^{(1D)}_{\rm
eff}= 7870 \pm 120\ $K and  $T^{(3D)}_{\rm eff}= 7890 \pm 120\ $K),
there is a possible solution for the model with $M_{\star}= 0.1650\
M_{\sun}$ at $T_{\rm  eff} \sim 7828\ $K. In Table~\ref{tab:perj2228}
we show the comparison between the observed and the theoretical
periods for this case.  In the case of $\ell=1, 2$, the absolute
maximum lies very close to the  ranges of allowed $T_{\rm  eff}$. It
corresponds to the model with $M_{\star}= 0.1554 \ M_{\sun}$ at
$T_{\rm  eff} \sim 7710$ K. In Table~\ref{tab:perj2228_b} we display
the comparison between the observed and the theoretical periods for
this model.

Since the solution for the case of $\ell= 1, 2$ ($M_{\star}= 0.1554\
M_{\sun}$ and $T_{\rm eff}= 7710\ $K)  implies a much better
period fit than the solution for the case of $\ell= 1$, since it lies
at a value of $T_{\rm  eff}$ almost compatible with the values given
by spectroscopy, and because its mass is in line with the
spectroscopic determinations for the mass ($M^{(1D)}_{\star}= 0.152\
M_{\sun}$ and $M^{(3D)}_{\star}= 0.142\ M_{\sun}$), we adopt this
model as the asteroseismological solution for J2228. According to our
nonadiabatic computations (Table~\ref{tab:perj2228_b}), most of the periods
of the adopted model are associated with pulsationally unstable modes.

\begin{table}[ht]
\centering
\caption{Same as Table~\ref{tab:perj1840_a} but for
J2228, for the model with $M_{\star}= 0.1650\ M_{\sun}$ and 
$T_{\rm eff}= 7828\ $K, in the case of $\ell= 1$. }
\begin{tabular}{ccccccc}
\hline
\hline
 $\Pi^{\rm O}$[s] & $\Pi^{\rm T}$[s] & $\ell$ & $k$ & $|\delta\Pi|$[s] & $\eta[10^{-8}]$ & 
Remark\\
\hline
$3254.5$& $3259.9$& 1 & $31$&  $5.4$ & $4.34$ & unstable \\
$4178.3$& $4175.8$& 1 & $40$&  $2.5$ & $5.03$ & unstable \\
$6234.9$& $6235.2$& 1 & $60$&  $0.3$ & $-7.02$ & stable \\
\hline
\hline
\end{tabular}
\label{tab:perj2228}
\end{table}

\begin{table}[ht]
\centering
\caption{Same as Table~\ref{tab:perj2228} but for the model adopted for J2228, with
$M_{\star}= 0.1554\ M_{\sun}$ and $T_{\rm eff}= 7710$ K,  in the case
of $\ell= 1,2$.} 
\begin{tabular}{cccccccc}
\hline
\hline
 $\Pi^{\rm O}$[s] & $\Pi^{\rm T}$[s] &  $\ell$ & $k$ & $|\delta\Pi|$[s] & $\eta[10^{-8}]$ & 
Remark\\
\noalign{\smallskip}
\hline
$3254.5$& $3254.2$& $2$& $52$&  $0.3$ & $1.61$ & unstable \\
$4178.3$& $4177.9$& $2$& $67$&  $0.4$ & $-2.67$ & stable \\
$6234.9$& $6234.4$& $1$& $58$&  $0.5$ & $0.832$ & unstable \\
\hline
\hline
\end{tabular}
\label{tab:perj2228_b}
\end{table}

\subsubsection{The case of J1738}
               
\begin{figure}  
\centering  
\includegraphics[clip,width=250pt]{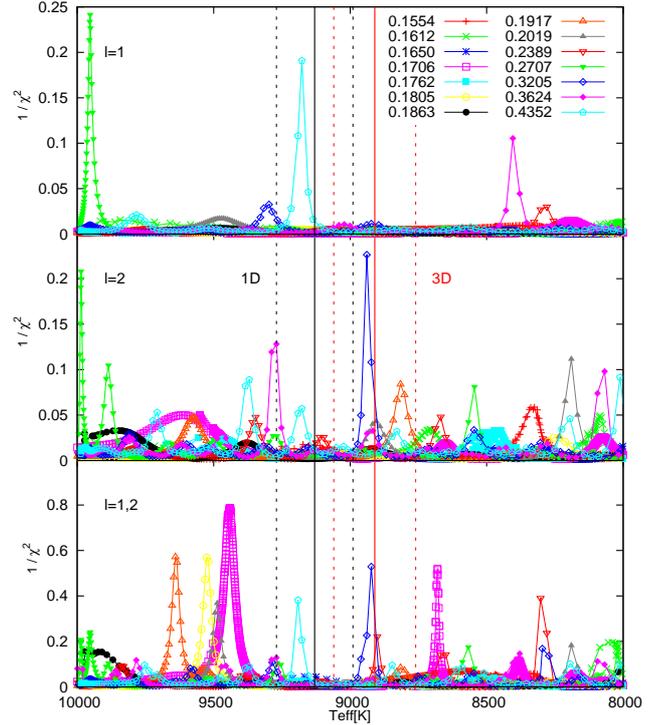}  
\caption{Same as Fig.~\ref{fig:ajustej1840}, but for the case of J1738.}
\label{fig:ajustej1738}
\end{figure}

In Fig.~\ref{fig:ajustej1738} we depict the match between the 
theoretical and the three observed periods of J1738, assuming they
are associated with $g$ modes, for the cases of $\ell= 1$ (top panel),
$\ell= 2$
(middle panel) and $\ell= 1, 2$ (bottom panel). In the case of $\ell= 1$,
the best solution lies at a very high value of $T_{\rm  eff}$, but
the second best solution lies within the range of allowed $T_{\rm  eff}$ (for the
1D atmosphere model determination, $T_{\rm eff}= 9130 \pm 140\ $K).
This solution is characterized by $M_{\star}= 0.4352\ M_{\sun}$ at
$T_{\rm  eff} \sim 9177\ $K, and the comparison between the observed and
the theoretical periods is shown in Table~\ref{tab:perj1738_a}. In the
case of $\ell= 1, 2$, the absolute maximum is located at a higher
effective temperature than the allowed by spectroscopy 
($T_{\rm eff}= 9130 \pm 140\ $K and $T_{\rm eff}= 8910 \pm 150\ $K, 1D and 3D models,
respectively), and there are many other
solutions. However, the models with $M_{\star}= 0.3205\ M_{\sun}$ at
$T_{\rm  eff} \sim 8922$ K and $M_{\star}= 0.4352\ M_{\sun}$ at
$T_{\rm  eff} \sim 9192$ K are relatively good period fits that lie within the
ranges of allowed $T_{\rm  eff}$. When we analyze in detail the period to period
fit, we see that the latter (shown in Table~\ref{tab:perj1738_b}) may be
more realistic due to the fact that more modes are associated with $\ell= 1$ than
$\ell= 2$. The opposite happens with the former, so we may rather choose the solution
with $M_{\star}= 0.4352\ M_{\sun}$ though is not the best one.

From this analysis, since the values of $(\chi^2)^{-1}$ for the two possible solutions are
not significantly different, and the value of the mass is the same for both of them, but
the solution for the $\ell= 1$ case has more periods associated with pulsationally unstable
modes, we conclude that this is the best asteroseismological solution, characterized by
$M_{\star}= 0.4352\ M_{\sun}$ and $T_{\rm  eff}=9177\ $K, which is in line with the $T_{\rm eff}$
given by the spectroscopy (for the 1D model atmosphere computations). However, when
we compare the mass of this model with the masses from the spectroscopic determinations
($M^{(1D)}_{\star}= 0.181\ M_{\sun}$ and $M^{(3D)}_{\star}= 0.172\ M_{\sun}$) we see that
they are not in good agreement. In summary, we cannot find any agreement
between the asteroseismological and the
spectroscopic results for J1738.

\begin{table}[ht]
\centering
\caption{Same as Table~\ref{tab:perj1840_a}, but for the model adopted for J1738, with
$M_{\star}= 0.4352\ M_{\sun}$ and $T_{\rm eff}= 9177\ $K, in the case of $\ell= 1$. }
\begin{tabular}{ccccccc}
\hline
\hline
 $\Pi^{\rm O}$[s] & $\Pi^{\rm T}$[s] & $\ell$ &$k$ & $|\delta\Pi|$[s] & $\eta[10^{-6}]$ & 
Remark\\
\hline
$1788$& $1788.9$& 1 &$30$&  $0.9$ & $3.41$ & unstable \\
$2656$& $2654.4$& 1 &$45$&  $1.6$ & $0.232$ & unstable \\
$3057$& $3060.5$& 1 &$52$&  $3.5$ & $-4.14$ & stable \\
\hline
\hline
\end{tabular}
\label{tab:perj1738_a}
\end{table}

\begin{table}[ht]
\centering
\caption{Same as Table~\ref{tab:perj1738_a}, for J1738 but for the model with
with $M_{\star}= 0.4352\ M_{\sun}$ and 
$T_{\rm eff}= 9192$ K, in the case of $\ell= 1, 2$. }
\begin{tabular}{cccccccc}
\hline
\hline
 $\Pi^{\rm O}$[s] & $\Pi^{\rm T}$[s]  &  $\ell$ & $k$ & $|\delta\Pi|$[s] & $\eta[10^{-6}]$ &
Remark\\
\noalign{\smallskip}
\hline
$1788$& $1786.7$& $1$& $30$&  $1.3$ & $4.45$ & unstable \\
$2656$& $2653.6$& $2$& $78$&  $2.4$ & $-31.0$ & stable \\
$3057$& $3056.5$& $1$& $52$&  $0.5$ & $-4.13$ & stable \\
\hline
\hline
\end{tabular}
\label{tab:perj1738_b}
\end{table}

\subsubsection{The case of J1618}

\begin{figure}  
\centering  
\includegraphics[clip,width=250pt]{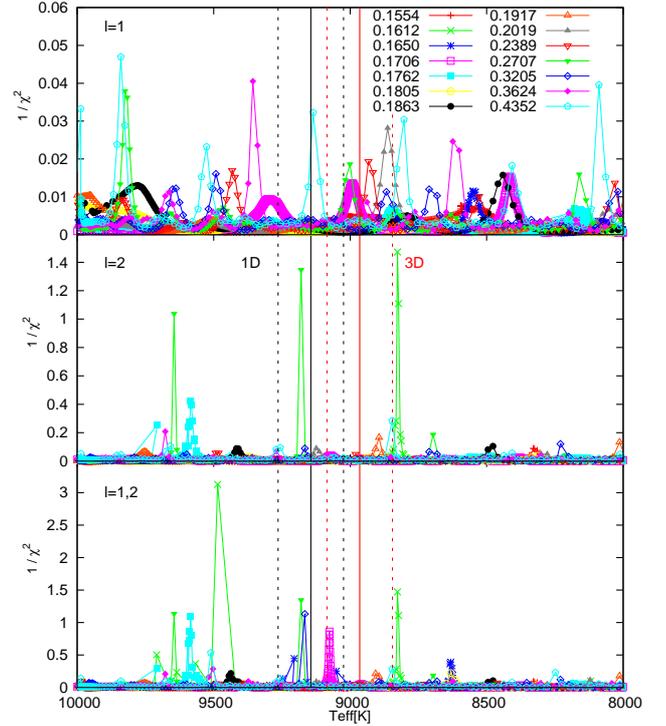}  
\caption{Same as Fig.~\ref{fig:ajustej1840}, but for the case of J1618.}
\label{fig:ajustej1618}
\end{figure}  

In Fig.~\ref{fig:ajustej1618} we show the match between the 
theoretical and the three observed periods of J1618, assuming they
are associated with $g$ modes, for the cases of $\ell= 1$ (top panel),
$\ell= 2$ (middle panel) and $\ell= 1, 2$ (bottom panel). In the case of
$\ell= 1$, we can see that there is no unambiguous solution. Within the range
of allowed $T_{\rm  eff}$ for the 1D model ($T_{\rm eff}= 9144 \pm 120\ $K), there is
a possible solution for the model with $M_{\star}= 0.4352\ M_{\sun}$ at
$T_{\rm  eff} \sim 9136\ $K, and
in the range of allowed $T_{\rm  eff}$ for the 3D model ($T_{\rm eff}= 8965 \pm 120\ $K)
there is another possible
solution for the model characterized by $M_{\star}= 0.2019\ M_{\sun}$ at
$T_{\rm  eff} \sim 8863\ $K. The latter may be more suitable as a solution
because its mass is in line with the spectroscopic determinations of the stellar
mass (and in comparison, the period fit for the other solution is not
significantly better). In Table~\ref{tab:perj1618_a} we show the comparison
between the observed and the theoretical periods for the model with
$M_{\star}= 0.2019\ M_{\sun}$.
The panel for the case of a mix of modes with $\ell= 1, 2$ shows an absolute
maximum for a higher value of $T_{\rm  eff}$ than allowed, and does not show any
unambiguous solution in the
ranges of allowed $T_{\rm  eff}$. However, there is a possible solution
characterized by $M_{\star}= 0.1706\ M_{\sun}$ at $T_{\rm  eff} \sim 9076$ K
because despite not being the best period fit in the ranges of
allowed $T_{\rm  eff}$, it corresponds to modes associated
both with $\ell= 1$ and $\ell= 2$ (and not only $\ell= 2$, see
Table~\ref{tab:perj1618_b}) and also the mass is quite in line with the
mass of the spectroscopic determination.

Although the solution for the $\ell= 1$ case ($M_{\star}= 0.2019\ M_{\sun}$)
has a mass slightly closer to the masses from the spectroscopic
determinations for J1618
($M^{(1D)}_{\star}= 0.220\ M_{\sun}$ and $M^{(3D)}_{\star}= 0.179\ M_{\sun}$) than the
solution for the $\ell= 1, 2$ case ($M_{\star}= 0.1706\ M_{\sun}$),
if we consider the fact that the latter is a better match between
the observed and the theoretical periods, and also that this value of the stellar
mass is more realistic for this type of stars, we conclude that the model with
$M_{\star}= 0.1706\ M_{\sun}$ and $T_{\rm eff}= 9076$ K, is a more suitable
solution. Also, the $T_{\rm eff}$ is in line with the spectroscopy.
Hence, this is the model we adopt for J1618. Note that all of the periods of the
adopted model are associated with pulsationally unstable modes.

\begin{table}[ht]
\centering
\caption{Same as Table~\ref{tab:perj1840_a}, but for J1618 for the model with
$M_{\star}= 0.2019\ M_{\sun}$ and $T_{\rm eff}= 8863\ $K, in the case of
$\ell= 1$. }
\begin{tabular}{ccccccc}
\hline
\hline
 $\Pi^{\rm O}$[s] & $\Pi^{\rm T}$[s] & $\ell$ & $k$ & $|\delta\Pi|$[s] & $\eta[10^{-6}]$ & 
Remark\\
\hline
$2543.0$ & $2546.46$ &  1 &$29$ &  $3.46$ & $1.70$ & unstable \\
$4935.21$& $4927.03$ &  1 &$57$ &  $8.18$ & $0.347$ & unstable \\
$6125.9$ & $6131.16$ &  1 &$71$ &  $5.26$ & $-4.11$ & stable \\
\hline
\hline
\end{tabular}
\label{tab:perj1618_a}
\end{table}

\begin{table}[ht]
\centering
\caption{Same as Table~\ref{tab:perj1618_a}, but for the model adopted for J1618
with $M_{\star}= 0.1706\ M_{\sun}$ and 
$T_{\rm eff}= 9076\ $K, in the case of $\ell= 1, 2$. }
\begin{tabular}{cccccccc}
\hline
\hline
 $\Pi^{\rm O}$[s] & $\Pi^{\rm T}$[s] &  $\ell$ & $k$ & $|\delta\Pi|$[s] & $\eta[10^{-4}]$ &
Remark\\
\noalign{\smallskip}
\hline
$2543.0$&  $2541.44$& $1$& $26$ &  $1.56$ & $0.0144$ & unstable \\
$4935.21$& $4934.59$& $2$& $91$ &  $0.62$ & $3.88$ & unstable \\
$6125.9$&  $6126.71$& $2$& $113$& $0.81$ & $1.32$ & unstable \\
\hline
\hline
\end{tabular}
\label{tab:perj1618_b}
\end{table}

\subsubsection{The case of J1735}

\begin{figure}  
\centering  
\includegraphics[clip,width=250pt]{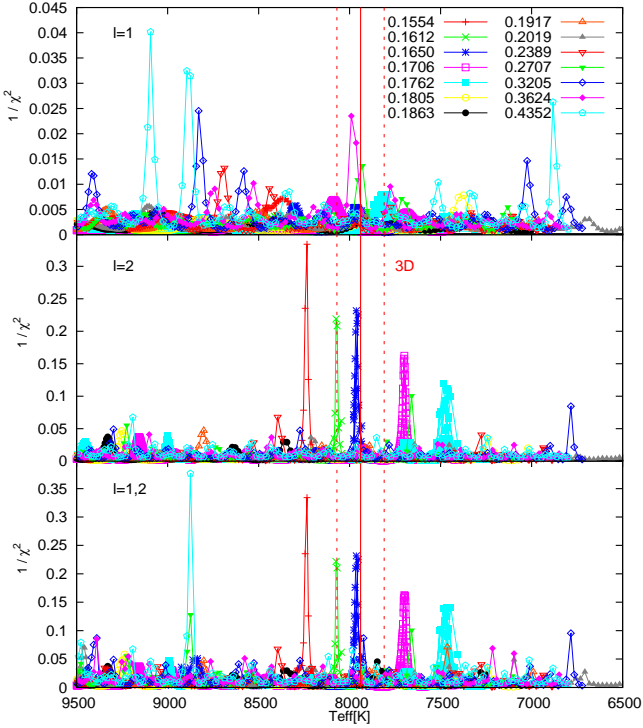}  
\caption{Same as Fig.~\ref{fig:ajustej1840}, but for the case of J1735.}
\label{fig:ajustej1735}
\end{figure}

In Fig.~\ref{fig:ajustej1735} we plot the match between the
theoretical and the four observed periods of J1735 assuming they are
associated with $g$ modes, for the cases of $\ell= 1$ (top panel),
$\ell= 2$ (middle panel) and $\ell= 1, 2$ (bottom panel). In the
$\ell= 1$ case, we can see that there are multiple local maxima that
have values of $T_{\rm  eff}$ which are either too high or too low in
comparison with the range of allowed $T_{\rm  eff}$ ($T_{\rm eff}=
7940 \pm 130\ $K).  However, there is a possible solution within that
range, corresponding to the model with $M_{\star}= 0.3624\ M_{\sun}$
at $T_{\rm  eff} \sim 7991\ $K. In Table~\ref{tab:perj1735_a} we show
the comparison between the observed and the theoretical periods for
the mentioned model. Note, however, that this stellar model that
constitutes a possible seismological solution for J1735, has all the modes
pulsationally stable. As for the case of $\ell= 1, 2$, we can see that
the best fit models have values of $T_{\rm  eff}$ higher than
allowed. Although there is a possible solution within the range of
allowed $T_{\rm  eff}$ for the model characterized by  $M_{\star}=
0.1650\ M_{\sun}$ at $T_{\rm  eff} \sim 7963\ $K, when we compare the
observed and the theoretical periods, we find that they are all
associated with $\ell= 2$. Then, the model with $M_{\star}= 0.1612\
M_{\sun}$ that lies at a slightly higher value of $T_{\rm  eff}$ than
allowed ($\sim 8075\ $K), may be a good solution for this case (see
Table~\ref{tab:perj1735_b}).

Taking into consideration the spectroscopic determination for the mass
of J1735, $M^{(3D)}_{\star}= 0.142 \pm 0.010 \ M_{\sun}$, and
comparing the quality of the period fit of the asteroseismological
results, we find that the model with $M_{\star}= 0.1612\ M_{\sun}$ and
$T_{\rm eff}= 8075\ $K, is an appropriate solution (with the
$T_{\rm eff}$ almost compatible with the spectroscopy) and this is the
one we adopt. This model has most of its periods associated with pulsationally
unstable modes.

\begin{table}[ht]
\centering
\caption{Same as Table~\ref{tab:perj1840_a}, but for J1735 for the model with
$M_{\star}= 0.3624\ M_{\sun}$ and $T_{\rm eff}= 7991\ $K, in
the case of $\ell= 1$. }
\begin{tabular}{ccccccc}
\hline
\hline
 $\Pi^{\rm O}$[s] & $\Pi^{\rm T}$[s] & $\ell$ & $k$ & $|\delta\Pi|$[s] & $\eta[10^{-6}]$ & 
Remark\\
\hline
$3362.76$& $3356.67$& 1 & $48$&  $6.09$ & $-0.660$ & stable \\
$3834.54$& $3841.23$& 1 & $55$&  $6.69$ & $-1.28$ & stable \\
$4541.88$& $4535.74$& 1 & $65$&  $6.14$ & $-2.73$ & stable \\
$4961.22$& $4954.12$& 1 & $71$&  $7.10$ & $-3.83$ & stable \\
\hline
\hline
\end{tabular}
\label{tab:perj1735_a}
\end{table}

\begin{table}[ht]
\centering
\caption{Same as Table~\ref{tab:perj1735_a}, but for the model adopted for J1735
with $M_{\star}= 0.1612\ M_{\sun}$ and 
$T_{\rm eff}= 8075\ $K, in the case of $\ell= 1, 2$. }
\begin{tabular}{cccccccc}
\hline
\hline
 $\Pi^{\rm O}$[s] & $\Pi^{\rm T}$[s] &  $\ell$ & $k$ & $|\delta\Pi|$[s] & $\eta[10^{-8}]$ &
Remark\\
\noalign{\smallskip}
\hline
$3362.76$& $3359.87$& $2$ & $56$&  $2.89$ & $5.57$ & unstable \\
$3834.54$& $3831.65$& $2$ & $64$&  $2.89$ & $0.243$ & unstable \\
$4541.88$& $4542.92$& $2$ & $76$&  $1.04$ & $-14.3$ & stable \\
$4961.22$& $4960.70$& $1$ & $48$&  $0.52$ & $13.9$ & unstable \\
\hline
\hline
\end{tabular}
\label{tab:perj1735_b}
\end{table}

\subsubsection{The case of J2139}

\begin{figure}  
\centering  
\includegraphics[clip,width=250pt]{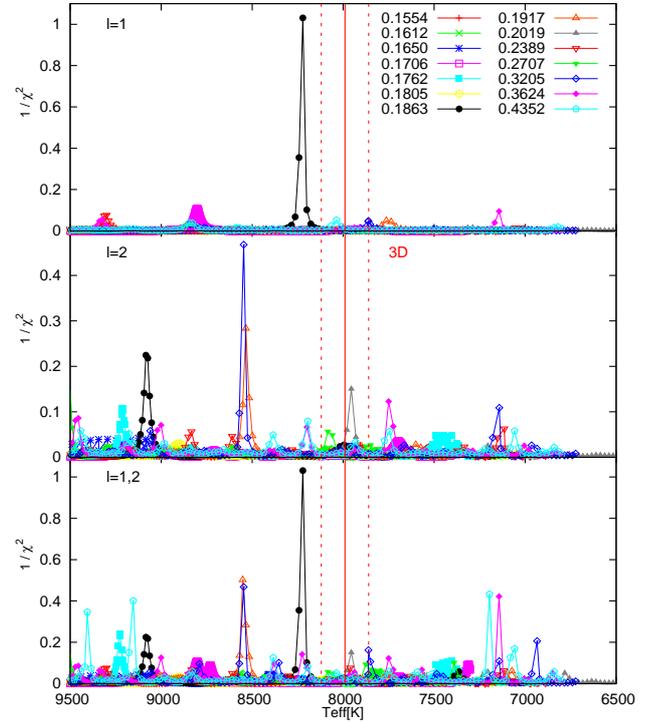}  
\caption{Same as Fig.~\ref{fig:ajustej1840}, but for the case of J2139.}
\label{fig:ajustej2139}
\end{figure}

In Fig.~\ref{fig:ajustej2139} we depict the match between the 
theoretical and the three observed periods of J2139 assuming they
are associated with $g$ modes, for the cases of $\ell= 1$ (top panel),
$\ell= 2$ (middle panel) and $\ell= 1, 2$ (bottom panel). In the first
case, we see that the absolute maximum, located at $T_{\rm eff} \sim 8221$ K
for a model with $M_{\star}= 0.1863\ M_{\sun}$, is very close to the
range of allowed $T_{\rm eff}$ ($T_{\rm eff}= 7990 \pm 130\ $K). Then, it represents
a good solution. In the
case of a mix of modes with $\ell= 1, 2$, the absolute maximum
corresponds to the same model, and the three periods are associated with
$\ell= 1$. There are other possible
solutions but they lie far from the range of allowed $T_{\rm eff}$, and inside
that range, the fits are not good. For the model with $M_{\star}= 0.1863\ M_{\sun}$,
we show the comparison between the observed and the theoretical periods in
Table~\ref{tab:perj2139}.

Considering these results, we may adopt the mentioned model,
with $M_{\star}= 0.1863\ M_{\sun}$ and $T_{\rm eff}= 8221\ $K,
which although is not quite in line with the spectroscopic result for the mass
($M^{(3D)}_{\star}= 0.149 \pm 0.011 \ M_{\sun}$) and the $T_{\rm eff}$,
is a very good period fit (with all the periods associated with $\ell= 1$ $g$ modes),
and also has all the periods associated with pulsationally unstable modes.

\begin{table}[ht]
\centering
\caption{Same as Table~\ref{tab:perj1840_a}, but for the model adopted for J2139
with $M_{\star}= 0.1863\ M_{\sun}$ and $T_{\rm eff}= 8221\ $K, in
the case of $\ell= 1$.}
\begin{tabular}{ccccccc}
\hline
\hline
 $\Pi^{\rm O}$[s] & $\Pi^{\rm T}$[s] & $\ell$ & $k$ & $|\delta\Pi|$[s] & $\eta[10^{-7}]$ & 
Remark\\
\hline
$2119.44$& $2120.01$& 1& $22$&  $0.57$ & $1.14$ & unstable \\
$2482.32$& $2483.89$& 1& $26$&  $1.57$ & $1.74$ & unstable \\
$3303.3$& $3303.63$&  1& $35$&  $0.33$ & $2.19$ & unstable \\
\hline
\hline
\end{tabular}
\label{tab:perj2139}
\end{table}

\section{Summary and conclusions}  
\label{conclusions}

\begin{table*}[t]
\centering
\caption{Stellar  masses (in solar units)  for all of the studied 
ELMV WD stars.}
\begin{tabular}{c|ccc|cc}
\hline 
\hline
Star &   & Period fit &      &\hspace{8mm} Spectroscopy   \\
\hline
    &  $ \ell= 1\ $ & $\ell= 1, 2\ $&  $\ell= 1,2 (g, p)$  & \hspace{8mm}(other works) &  \\
    &    $  (g) $     & $(g)$         &  $\ell= 0$ (radial)  & 1D & 3D\\
\hline
J1840   & 0.2389         & 0.1805             & ---               & 0.183$^{\rm a}$   & 0.177$^{\rm f}$ \\
J1112   & 0.3205$^{\rm *}$  & 0.2389$^{\rm *}$   & 0.1612$^{\rm **}$   & 0.179$^{\rm b}$  & 0.169$^{\rm f}$ \\
J1518   & 0.3205         & 0.2707             & ---               & 0.220$^{\rm b}$   & 0.197$^{\rm f}$ \\ 
J1614   & 0.1762         & 0.3205             & ---              & 0.192$^{\rm c}$   & 0.172$^{\rm f}$ \\
J2228   & 0.1650         & 0.1554           & ---               & 0.152$^{\rm c}$   & 0.142$^{\rm f}$ \\
J1738   & 0.4352         & 0.4352           & ---               & 0.181$^{\rm d}$   & 0.172$^{\rm f}$ \\
J1618   & 0.2019         & 0.1706           & ---             & 0.220$^{\rm e}$   & 0.179$^{\rm f}$ \\
J1735   & 0.3624         & 0.1612           & ---             &   ---           & 0.142$^{\rm g}$ \\
J2139   & 0.1863         &  ---             &   ---           &   ---           & 0.149$^{\rm g}$ \\
\hline
\hline
\end{tabular} 
\label{tabla-masas}

{\footnotesize  Notes: 
$^{\rm a}$\citet{2012ApJ...750L..28H}.
$^{\rm *}$ Determined using a subset of the observed periods.
$^{\rm **}$ Determined using the whole set of the observed periods.
$^{\rm b}$\citet{2013ApJ...765..102H}.  
$^{\rm c}$\citet{2013MNRAS.436.3573H}.  
$^{\rm d}$\citet{2015MNRAS.446L..26K}.  
$^{\rm e}$\citet{2015ASPC..493..217B}.  
$^{\rm f}$ Determined using the corrections for 3D effects by \citet{2015ApJ...809..148T}.}
$^{\rm g}$\citet{2017ApJ...835..180B}.
\end{table*}

In  this  work,  we  have presented  a detailed asteroseismological   study
of all the known pulsating ELM WD stars (ELMVs), considering the pulsation
spectrum they exhibit and employing the set of evolutionary models of
\cite{2013A&A...557A..19A}. This is the fifth paper in a series of works
dedicated on pulsating low-mass, He-core WDs (including ELMV WDs).
The present paper is devoted to perform the first asteroseismological analysis
of all the known ELMV stars. For this purpose we employed some asteroseismological
tools. One of them is based on the comparison between the observed period
spacing of the star under analysis with the average of the period
spacings computed on our grid of models. So firstly, we tried to determine
the observed period spacing for each target star, through three independent
significance tests. Given that the stars under study exhibit few periods,
we could only follow this approach for the cases of the stars showing four
periods or more, i.e. J1840, J1112, J1518 and J1735. However, for the first
two stars we could not find any unambiguous constant period spacing. In
the case of J1518 and J1735, on the other hand, we found a clear indication
of a constant period spacing at roughly $44\ $s and $59\ $s, respectively,
from the three significance tests applied. After comparing these values
with the average of the computed period spacings for our grid of models,
we found that the resulting stellar masses (greater than
$0.4352\ M_{\sun}$ in both cases) are higher than expected for this type of stars.
In the case of J1518, it may be associated with the fact that this star is
not pulsating in the asymptotic regime \citep{2014A&A...569A.106C}.
The case of J1735 is more intriguing because this star seems to be in that regime.

Next, we searched for the best-fit model, i.e. the theoretical
model that provides the best match between the
individual pulsation periods exhibited by the star  and the theoretical
pulsation periods. We assessed the function
$\chi^2=\chi^2(M_{\star}, T_{\rm eff})$  (given by
Eq.~\ref{chi} of Section~\ref{fitting}) for our complete set of model sequences,
covering a wide range in effective temperature 
 ($13000 \gtrsim T_{\rm eff} \gtrsim 6000\ $ K). Due to the multiplicity of
 solutions, we were forced to employ some external constraints (for instance, the
 uncertainty in the $T_{\rm eff}$, given by spectroscopy).
 We assumed that all of the observed periods correspond to $\ell = 1$ g modes
 and considered them to compute the quality function for each target star.
We also considered the (unlikely) case in which all of the
observed periods correspond to $\ell = 2$ g modes. Finally, we considered the
case of a mix of $\ell= 1$ and $\ell= 2$ g modes. For the particular case of the
star J1112, we performed two different analyses. Since the two shortest periods
reported for this star are not confirmed \citep{2013ApJ...765..102H}, first we
carried out a period fit applied to the subset of the five longest periods exhibited by this
star considering they are associated with $\ell= 1$, $\ell= 2$ and a mix of
$\ell= 1$ and $\ell= 2$ g modes. Second, for the whole set of seven periods, we explored two
possibilities: that all of the observed periods correspond to a mix of $g$
and $p$ modes ($\ell= 1$), and also the case in which the observed periods
correspond to radial ($\ell= 0$) and $p$ and $g$ modes ($\ell= 1, 2$). In
Table~\ref{tabla-masas} we show a compilation of the mass determinations for
the ELMVs both from spectroscopic (other works) and period-fit results
(this work). Considering the obtained results, we found that the
seismological mass is in good agreement with the spectroscopic determinations
for J1840 (in the case of a mix of $\ell= 1, 2$ g modes), J1614
(for the case of $\ell= 1$ $g$ modes), J2228 (for the case of $\ell= 1, 2$ $g$
modes), J1618 (for the case of $\ell= 1, 2$ $g$ modes), and
J1735 (for the case of $\ell= 1, 2$ $g$ modes). We consider that there is a
good agreement between the seismological and spectroscopic mass when the difference
is below the uncertainty of 15\%, that is, the typical difference in the mass value
derived from independent sets of evolutionary tracks. Then, we conclude that for most
of the target stars, the adopted models
from the asteroseismological analyses have masses which are
in line with the spectroscopic results. At variance with this, for four stars
(J1738, J1518, J1112 and J2139) we obtained a larger value of the stellar mass
in comparison with the spectroscopic determinations. In particular,
gathering together the mass determinations for J1518, we conclude that there is
no agreement between the mass given by the spectroscopy
($M^{1D}_{\star}= 0.220\ M_{\sun}$  and $M^{3D}_{\star}= 0.197\ M_{\sun}$),
the mass obtained from the comparison between the
observed period spacing and the average of the computed period spacings (which
is larger than $0.4352\ M_{\sun}$), and the mass from the adopted asteroseismological
model ($M_{\star}= 0.2707\ M_{\sun}$).
We also mention that, in spite of the fact that we were able to adopt a
seismological model for J1735 whose mass ($M_{\star}= 0.1612\ M_{\sun}$) is in line
with the spectroscopic determination ($M_{\star}= 0.142 \pm 0.010\ M_{\sun}$), we could
not find such agreement for the mass resulting from the comparison between the
observed period spacing and the average of the computed period spacings
($M_{\star} \gtrsim 0.44\ M_{\sun}$). Reversing the argument, as this star seems to
be in the asymptotic regime, if the value we have obtained for the period spacing of
this star were in fact associated with high-radial order $g$ modes, it could indicate
that the stellar mass is higher than the determined by the spectroscopic and
the period-to-period fit analysis, though in that case the star could not be classified
as an ELMV WD star. Finally, it is worth mentioning that, in general, the pulsation periods
corresponding to the asteroseismological models adopted in this work for the analysed
ELMV WDs are pulsationally unstable, according to our nonadiabatic computations.
This agreement between the adiabatic and nonadiabatic predictions gives
more relevance to our asteroseismological results.

From the results presented in this paper for all the known ELMVs, it is evident
once again the power of this approach, since in most of the cases we were able
to constrain the value of the stellar mass. Moreover, once a model has been
adopted we can access to additional information, as can be seen in
Table~\ref{tab:tablafinal}, which is another advantage of asteroseismology.
Taking into account these results, four of the stars under analysis (J1840, J1518,
J1738 and J2139) are not strictly ELM WD according to our definition previously
stated, that is, the progenitors of these stars might have experienced multiple
flashes.

\begin{figure}
\begin{center}
\includegraphics[clip,width=9 cm]{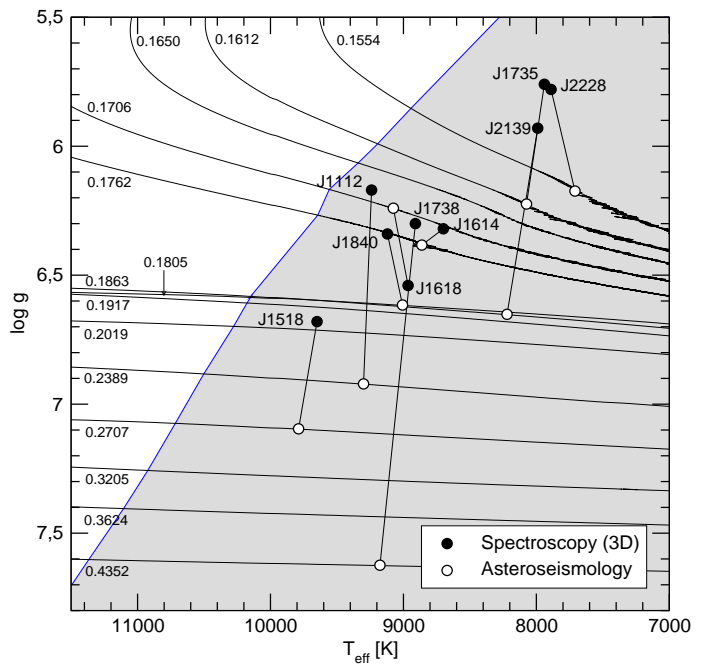} 
\caption{The location of the nine analyzed stars, according to the spectroscopic parameters
from the 3D model-atmosphere (black circles), and the corresponding
asteroseismological models adopted for each star (white circles), along with our
evolutionary tracks of low mass He-core WDs in the $\log T_{\rm eff}-\log g$ plane.
The gray-shaded region bounded by the thin blue line corresponds to the instability
domain of $\ell= 1$  $g$ modes modes according to the nonadiabatic computations
of \citet{2016A&A...585A...1C}.}
\label{HR-comparacion} 
\end{center}
\end{figure}

In Fig.~\ref{HR-comparacion} we show the location of the
nine analysed ELMVs (according to the 3D model-atmosphere parameters)
and the corresponding values of $T_{\rm eff}$ and $\log g$ of the
asteroseismological models adopted for each star,
along with our evolutionary tracks of low mass He-core WDs and the
the instability domain of $\ell= 1$  $g$ modes computed by
\citet{2016A&A...585A...1C}.  As we mentioned, for five stars we
found good agreement between the seismological mass and the spectroscopic one,
and for the remaining four stars the agreement is not good.
Beyond that, Fig.~\ref{HR-comparacion} demonstrates  that
for eight out of nine stars analyzed, the asteroseismological models are more
massive (i.e., they are characterized by higher gravities) in comparison with the
spectroscopy results. This systematic trend is also found in the case (not shown)
in which the $T_{\rm eff}$ and $\log(g)$ values derived from calculations of 1D
atmospheres are adopted. This trend could be related, in part, to the fact that we are not
considering low-mass He-core WD models characterized by outer H envelopes thinner
than those predicted by the complete binary evolutionary history of the progenitor
stars. Alternatively, it could be an indication that the
spectroscopic determinations of $\log g$ and $T_{\rm eff}$ in this class of stars
were not proper. 

\begin{table*}[ t]
\centering
\caption{Main characteristics of the adopted 
asteroseismological model for every known ELMV WD.}
\begin{tabular}{cccccc}
\hline
\hline
Star & $T_{\rm eff}$ [K]  & $\log(g)$ [cgs]  &  $M_{\star}[M_{\sun}]$& $\log(R_{\star}/R_{\sun})$ & $\log(L_{\star}/L_{\sun})$ \\
\hline
J1840             & 9007  &     6.6156     & 0.1805     & -1.4609    &  -2.1487 \\
J1112$^{\rm *}$     & 9300  &     6.9215     & 0.2389     & -1.5528    &  -2.2757 \\
J1518$^{\rm *}$     & 9789 &      7.0956     & 0.2707     & -1.6126    &  -2.3098 \\
J1614             & 8862  &     6.3832     & 0.1762     & -1.3497    &  -1.9547 \\
J2228             & 7710  &     6.1738     & 0.1554     & -1.2725    &  -2.0409 \\
J1738$^{\rm *}$     & 9177  &     7.6241     & 0.4352     & -1.7746    &  -2.7447 \\
J1618             & 9076  &     6.2403     & 0.1706     & -1.2857    &  -1.7852 \\
J1735             & 8075  &     6.2241     & 0.1612     & -1.2899    &  -1.9957 \\
J2139$^{\rm *}$    & 8221  &     6.6515     & 0.1863     & -1.4724     &  -2.3279 \\
\hline 
\hline
\end{tabular}
\label{tab:tablafinal}

{\footnotesize
Note: $^{\rm *}$ Solution whose mass is in conflict with the spectroscopic results.}

\end{table*}

In this paper, we have considered low-mass He-core WDs coming from solar
metallicity progenitors, typical of the population of the Galactic Disk. The threshold
in the stellar mass value, below which CNO flashes on the early WD cooling branch are
not expected to occur, is $\sim 0.18 M_{\sun}$. If we had adopted
progenitors with lower metallicities, representative of the population of the Galactic
Halo, the threshold mass limit should be larger (see
\cite{2002MNRAS.337.1091S,2004ApJ...616.1124N,2016A&A...595A..35I}. Also, the H envelope
of the low-mass WDs should be
thicker than those obtained in \cite{2013A&A...557A..19A} \citep[e.g.][]{2016A&A...595A..35I}.
So, by assuming 
that some of the ELMVs studied in this work are actually objects of the Galactic Halo,
the asteroseismological analysis should be based on evolutionary models coming from low
metallicity progenitors, and therefore the characteristics of the asteroseismological
models for each analyzed star could be different to those obtained in this work.       

In this study, we have considered low-mass He-core WD models characterized
by thick outer H envelopes, consistent with the previous evolution.
We are well aware that there are strong uncertainties about the precise value
of the thickness of this envelope. We cannot discard that WD models with H
envelopes thinner than those characterizing our set of models could result
from binary evolution computations that assume different angular-momentum loss
prescriptions due to mass loss, different initial mass-ratio, etc, than that adopted
in \cite{2013A&A...557A..19A} \citep[see the detailed works by][]{2015ASPC..493..487I,2016A&A...595A..35I}.
Asteroseismological analyses considering
low-mass He-core WD models characterized by thinner outer H envelopes will
be the core feature of a future work.

%------------------------------------------------------------------------  
  
\begin{acknowledgements}
We wish to thank our anonymous referee for the constructive
comments and suggestions that greatly improved the original version of
the paper.
Part of this work was supported by AGENCIA through the 
Programa de Modernizaci\'on Tecnol\'ogica BID 1728/OC-AR, and by the 
PIP 112-200801-00940 grant from CONICET. This research made use 
of NASA Astrophysics Data System.
\end{acknowledgements}

\bibliographystyle{aa}
\bibliography{biblio}

\begin{thebibliography}{88}
\expandafter\ifx\csname natexlab\endcsname\relax\def\natexlab#1{#1}\fi

\bibitem[{{Althaus} {et~al.}(2015){Althaus}, {Camisassa}, {Miller Bertolami},
  {C{\'o}rsico}, \& {Garc{\'{\i}}a-Berro}}]{2015A&A...576A...9A}
{Althaus}, L.~G., {Camisassa}, M.~E., {Miller Bertolami}, M.~M., {C{\'o}rsico},
  A.~H., \& {Garc{\'{\i}}a-Berro}, E. 2015, \aap, 576, A9

\bibitem[{{Althaus} {et~al.}(2010){Althaus}, {C{\'o}rsico}, {Isern}, \&
  {Garc{\'{\i}}a-Berro}}]{review}
{Althaus}, L.~G., {C{\'o}rsico}, A.~H., {Isern}, J., \& {Garc{\'{\i}}a-Berro},
  E. 2010, \aapr, 18, 471

\bibitem[{{Althaus} {et~al.}(2013){Althaus}, {Miller Bertolami}, \&
  {C{\'o}rsico}}]{2013A&A...557A..19A}
{Althaus}, L.~G., {Miller Bertolami}, M.~M., \& {C{\'o}rsico}, A.~H. 2013,
  \aap, 557, A19

\bibitem[{{Althaus} {et~al.}(2009){Althaus}, {Panei}, {Romero}, {Rohrmann},
  {C{\'o}rsico}, {Garc{\'{\i}}a-Berro}, \& {Miller
  Bertolami}}]{2009A&A...502..207A}
{Althaus}, L.~G., {Panei}, J.~A., {Romero}, A.~D., {et~al.} 2009, \aap, 502,
  207

\bibitem[{{Althaus} {et~al.}(2005){Althaus}, {Serenelli}, {Panei},
  {C{\'o}rsico}, {Garc{\'{\i}}a-Berro}, \&
  {Sc{\'o}ccola}}]{2005A&A...435..631A}
{Althaus}, L.~G., {Serenelli}, A.~M., {Panei}, J.~A., {et~al.} 2005, \aap, 435,
  631

\bibitem[{{Bell} {et~al.}(2017){Bell}, {Gianninas}, {Hermes}, {Winget},
  {Kilic}, {Montgomery}, {Castanheira}, {Vanderbosch}, {Winget}, \&
  {Brown}}]{2017ApJ...835..180B}
{Bell}, K.~J., {Gianninas}, A., {Hermes}, J.~J., {et~al.} 2017, \apj, 835, 180

\bibitem[{{Bell} {et~al.}(2015){Bell}, {Kepler}, {Montgomery}, {Hermes},
  {Harrold}, \& {Winget}}]{2015ASPC..493..217B}
{Bell}, K.~J., {Kepler}, S.~O., {Montgomery}, M.~H., {et~al.} 2015, in
  Astronomical Society of the Pacific Conference Series, Vol. 493, 19th
  European Workshop on White Dwarfs, ed. P.~{Dufour}, P.~{Bergeron}, \&
  G.~{Fontaine}, 217

\bibitem[{{Bischoff-Kim} {et~al.}(2008){Bischoff-Kim}, {Montgomery}, \&
  {Winget}}]{2008ApJ...675.1505B}
{Bischoff-Kim}, A., {Montgomery}, M.~H., \& {Winget}, D.~E. 2008, \apj, 675,
  1505

\bibitem[{{Bogn{\'a}r} {et~al.}(2014){Bogn{\'a}r}, {Papar{\'o}}, {C{\'o}rsico},
  {Kepler}, \& {Gy{\H o}rffy}}]{2014A&A...570A.116B}
{Bogn{\'a}r}, Z., {Papar{\'o}}, M., {C{\'o}rsico}, A.~H., {Kepler}, S.~O., \&
  {Gy{\H o}rffy}, {\'A}. 2014, \aap, 570, A116

\bibitem[{{Bogn{\'a}r} {et~al.}(2016){Bogn{\'a}r}, {Papar{\'o}}, {Moln{\'a}r},
  {P{\'a}pics}, {Plachy}, {Vereb{\'e}lyi}, \&
  {S{\'o}dor}}]{2016MNRAS.461.4059B}
{Bogn{\'a}r}, Z., {Papar{\'o}}, M., {Moln{\'a}r}, L., {et~al.} 2016, \mnras,
  461, 4059

\bibitem[{{Bradley}(1998)}]{1998ApJS..116..307B}
{Bradley}, P.~A. 1998, \apjs, 116, 307

\bibitem[{{Bradley}(2001)}]{2001ApJ...552..326B}
{Bradley}, P.~A. 2001, \apj, 552, 326

\bibitem[{{Brassard} {et~al.}(1991){Brassard}, {Fontaine}, {Wesemael},
  {Kawaler}, \& {Tassoul}}]{1991ApJ...367..601B}
{Brassard}, P., {Fontaine}, G., {Wesemael}, F., {Kawaler}, S.~D., \& {Tassoul},
  M. 1991, \apj, 367, 601

\bibitem[{{Brickhill}(1991)}]{1991MNRAS.251..673B}
{Brickhill}, A.~J. 1991, \mnras, 251, 673

\bibitem[{{Brown} {et~al.}(2016){Brown}, {Gianninas}, {Kilic}, {Kenyon}, \&
  {Allende Prieto}}]{2016ApJ...818..155B}
{Brown}, W.~R., {Gianninas}, A., {Kilic}, M., {Kenyon}, S.~J., \& {Allende
  Prieto}, C. 2016, \apj, 818, 155

\bibitem[{{Brown} {et~al.}(2013){Brown}, {Kilic}, {Allende Prieto},
  {Gianninas}, \& {Kenyon}}]{2013ApJ...769...66B}
{Brown}, W.~R., {Kilic}, M., {Allende Prieto}, C., {Gianninas}, A., \&
  {Kenyon}, S.~J. 2013, \apj, 769, 66

\bibitem[{{Brown} {et~al.}(2010){Brown}, {Kilic}, {Allende Prieto}, \&
  {Kenyon}}]{2010ApJ...723.1072B}
{Brown}, W.~R., {Kilic}, M., {Allende Prieto}, C., \& {Kenyon}, S.~J. 2010,
  \apj, 723, 1072

\bibitem[{{Brown} {et~al.}(2012){Brown}, {Kilic}, {Allende Prieto}, \&
  {Kenyon}}]{2012ApJ...744..142B}
{Brown}, W.~R., {Kilic}, M., {Allende Prieto}, C., \& {Kenyon}, S.~J. 2012,
  \apj, 744, 142

\bibitem[{{Brown} {et~al.}(2017){Brown}, {Kilic}, \&
  {Gianninas}}]{2017ApJ...839...23B}
{Brown}, W.~R., {Kilic}, M., \& {Gianninas}, A. 2017, \apj, 839, 23

\bibitem[{{Burgers}(1969)}]{1969fecg.book.....B}
{Burgers}, J.~M. 1969, {Flow Equations for Composite Gases} (New York: Academic
  Press)

\bibitem[{{Calcaferro} {et~al.}(2016){Calcaferro}, {C{\'o}rsico}, \&
  {Althaus}}]{2016A&A...589A..40C}
{Calcaferro}, L.~M., {C{\'o}rsico}, A.~H., \& {Althaus}, L.~G. 2016, \aap, 589,
  A40

\bibitem[{{Calcaferro} {et~al.}(2017){Calcaferro}, {C{\'o}rsico}, \&
  {Althaus}}]{2017A&A...600A..73C}
{Calcaferro}, L.~M., {C{\'o}rsico}, A.~H., \& {Althaus}, L.~G. 2017, \aap, 600,
  A73

\bibitem[{{Cassisi} {et~al.}(2007){Cassisi}, {Potekhin}, {Pietrinferni},
  {Catelan}, \& {Salaris}}]{2007ApJ...661.1094C}
{Cassisi}, S., {Potekhin}, A.~Y., {Pietrinferni}, A., {Catelan}, M., \&
  {Salaris}, M. 2007, \apj, 661, 1094

\bibitem[{{Castanheira} \& {Kepler}(2008)}]{2008MNRAS.385..430C}
{Castanheira}, B.~G. \& {Kepler}, S.~O. 2008, \mnras, 385, 430

\bibitem[{{Castanheira} \& {Kepler}(2009)}]{2009MNRAS.396.1709C}
{Castanheira}, B.~G. \& {Kepler}, S.~O. 2009, \mnras, 396, 1709

\bibitem[{{C{\'o}rsico} \& {Althaus}(2006)}]{2006A&A...454..863C}
{C{\'o}rsico}, A.~H. \& {Althaus}, L.~G. 2006, \aap, 454, 863

\bibitem[{{C{\'o}rsico} \& {Althaus}(2014{\natexlab{a}})}]{2014A&A...569A.106C}
{C{\'o}rsico}, A.~H. \& {Althaus}, L.~G. 2014{\natexlab{a}}, \aap, 569, A106

\bibitem[{{C{\'o}rsico} \& {Althaus}(2014{\natexlab{b}})}]{2014ApJ...793L..17C}
{C{\'o}rsico}, A.~H. \& {Althaus}, L.~G. 2014{\natexlab{b}}, \apjl, 793, L17

\bibitem[{{C{\'o}rsico} \& {Althaus}(2016)}]{2016A&A...585A...1C}
{C{\'o}rsico}, A.~H. \& {Althaus}, L.~G. 2016, \aap, 585, A1

\bibitem[{{C{\'o}rsico} {et~al.}(2008){C{\'o}rsico}, {Althaus}, {Kepler},
  {Costa}, \& {Miller Bertolami}}]{2008A&A...478..869C}
{C{\'o}rsico}, A.~H., {Althaus}, L.~G., {Kepler}, S.~O., {Costa}, J.~E.~S., \&
  {Miller Bertolami}, M.~M. 2008, \aap, 478, 869

\bibitem[{{C{\'o}rsico} {et~al.}(2006){C{\'o}rsico}, {Althaus}, \& {Miller
  Bertolami}}]{2006A&A...458..259C}
{C{\'o}rsico}, A.~H., {Althaus}, L.~G., \& {Miller Bertolami}, M.~M. 2006,
  \aap, 458, 259

\bibitem[{{C{\'o}rsico} {et~al.}(2012{\natexlab{a}}){C{\'o}rsico}, {Althaus},
  {Miller Bertolami}, \& {Bischoff-Kim}}]{2012A&A...541A..42C}
{C{\'o}rsico}, A.~H., {Althaus}, L.~G., {Miller Bertolami}, M.~M., \&
  {Bischoff-Kim}, A. 2012{\natexlab{a}}, \aap, 541, A42

\bibitem[{{C{\'o}rsico} {et~al.}(2009){C{\'o}rsico}, {Althaus}, {Miller
  Bertolami}, \& {Garc{\'{\i}}a-Berro}}]{2009A&A...499..257C}
{C{\'o}rsico}, A.~H., {Althaus}, L.~G., {Miller Bertolami}, M.~M., \&
  {Garc{\'{\i}}a-Berro}, E. 2009, \aap, 499, 257

\bibitem[{{C{\'o}rsico} {et~al.}(2007{\natexlab{a}}){C{\'o}rsico}, {Althaus},
  {Miller Bertolami}, \& {Werner}}]{2007A&A...461.1095C}
{C{\'o}rsico}, A.~H., {Althaus}, L.~G., {Miller Bertolami}, M.~M., \& {Werner},
  K. 2007{\natexlab{a}}, \aap, 461, 1095

\bibitem[{{C{\'o}rsico} {et~al.}(2016){C{\'o}rsico}, {Althaus}, {Serenelli},
  {Kepler}, {Jeffery}, \& {Corti}}]{2016A&A...588A..74C}
{C{\'o}rsico}, A.~H., {Althaus}, L.~G., {Serenelli}, A.~M., {et~al.} 2016,
  \aap, 588, A74

\bibitem[{{C{\'o}rsico} {et~al.}(2007{\natexlab{b}}){C{\'o}rsico}, {Miller
  Bertolami}, {Althaus}, {Vauclair}, \& {Werner}}]{2007A&A...475..619C}
{C{\'o}rsico}, A.~H., {Miller Bertolami}, M.~M., {Althaus}, L.~G., {Vauclair},
  G., \& {Werner}, K. 2007{\natexlab{b}}, \aap, 475, 619

\bibitem[{{C{\'o}rsico} {et~al.}(2012{\natexlab{b}}){C{\'o}rsico}, {Romero},
  {Althaus}, \& {Hermes}}]{2012A&A...547A..96C}
{C{\'o}rsico}, A.~H., {Romero}, A.~D., {Althaus}, L.~G., \& {Hermes}, J.~J.
  2012{\natexlab{b}}, \aap, 547, A96

\bibitem[{{Corti} {et~al.}(2016){Corti}, {Kanaan}, {C{\'o}rsico}, {Kepler},
  {Althaus}, {Koester}, \& {S{\'a}nchez Arias}}]{2016A&A...587L...5C}
{Corti}, M.~A., {Kanaan}, A., {C{\'o}rsico}, A.~H., {et~al.} 2016, \aap, 587,
  L5

\bibitem[{{Dziembowski}(1977)}]{1977AcA....27..203D}
{Dziembowski}, W. 1977, \actaa, 27, 203

\bibitem[{{Dziembowski}(1971)}]{1971AcA....21..289D}
{Dziembowski}, W.~A. 1971, \actaa, 21, 289

\bibitem[{{Fontaine} \& {Brassard}(2008)}]{2008PASP..120.1043F}
{Fontaine}, G. \& {Brassard}, P. 2008, \pasp, 120, 1043

\bibitem[{{Fontaine} {et~al.}(2017){Fontaine}, {Istrate}, {Gianninas},
  {Brassard}, \& {Van Grootel}}]{2017ASPC..509..347F}
{Fontaine}, G., {Istrate}, A., {Gianninas}, A., {Brassard}, P., \& {Van
  Grootel}, V. 2017, in Astronomical Society of the Pacific Conference Series,
  Vol. 509, 20th European White Dwarf Workshop, ed. P.-E. {Tremblay},
  B.~{Gaensicke}, \& T.~{Marsh}, 347

\bibitem[{{Giammichele} {et~al.}(2017{\natexlab{a}}){Giammichele}, {Charpinet},
  {Brassard}, \& {Fontaine}}]{2017A&A...598A.109G}
{Giammichele}, N., {Charpinet}, S., {Brassard}, P., \& {Fontaine}, G.
  2017{\natexlab{a}}, \aap, 598, A109

\bibitem[{{Giammichele} {et~al.}(2017{\natexlab{b}}){Giammichele}, {Charpinet},
  {Fontaine}, \& {Brassard}}]{2017ApJ...834..136G}
{Giammichele}, N., {Charpinet}, S., {Fontaine}, G., \& {Brassard}, P.
  2017{\natexlab{b}}, \apj, 834, 136

\bibitem[{{Giammichele} {et~al.}(2016){Giammichele}, {Fontaine}, {Brassard}, \&
  {Charpinet}}]{2016ApJS..223...10G}
{Giammichele}, N., {Fontaine}, G., {Brassard}, P., \& {Charpinet}, S. 2016,
  \apjs, 223, 10

\bibitem[{{Gianninas} {et~al.}(2016){Gianninas}, {Curd}, {Fontaine}, {Brown},
  \& {Kilic}}]{2016ApJ...822L..27G}
{Gianninas}, A., {Curd}, B., {Fontaine}, G., {Brown}, W.~R., \& {Kilic}, M.
  2016, \apjl, 822, L27

\bibitem[{{Gianninas} {et~al.}(2014{\natexlab{a}}){Gianninas}, {Dufour},
  {Kilic}, {Brown}, {Bergeron}, \& {Hermes}}]{2014ApJ...794...35G}
{Gianninas}, A., {Dufour}, P., {Kilic}, M., {et~al.} 2014{\natexlab{a}}, \apj,
  794, 35

\bibitem[{{Gianninas} {et~al.}(2014{\natexlab{b}}){Gianninas}, {Hermes},
  {Brown}, {Dufour}, {Barber}, {Kilic}, {Kenyon}, \&
  {Harrold}}]{2014ApJ...781..104G}
{Gianninas}, A., {Hermes}, J.~J., {Brown}, W.~R., {et~al.} 2014{\natexlab{b}},
  \apj, 781, 104

\bibitem[{{Gianninas} {et~al.}(2015){Gianninas}, {Kilic}, {Brown}, {Canton}, \&
  {Kenyon}}]{2015ApJ...812..167G}
{Gianninas}, A., {Kilic}, M., {Brown}, W.~R., {Canton}, P., \& {Kenyon}, S.~J.
  2015, \apj, 812, 167

\bibitem[{{Haft} {et~al.}(1994){Haft}, {Raffelt}, \&
  {Weiss}}]{1994ApJ...425..222H}
{Haft}, M., {Raffelt}, G., \& {Weiss}, A. 1994, \apj, 425, 222

\bibitem[{{Handler} {et~al.}(1997){Handler}, {Pikall}, {O'Donoghue}, {Buckley},
  {Vauclair}, {Chevreton}, {Giovannini}, {Kepler}, {Goode}, {Provencal},
  {Wood}, {Clemens}, {O'Brien}, {Nather}, {Winget}, {Kleinman}, {Kanaan},
  {Watson}, {Nitta}, {Montgomery}, {Klumpe}, {Bradley}, {Sullivan}, {Wu},
  {Marar}, {Seetha}, {Ashoka}, {Mahra}, {Bhat}, {Babu}, {Leibowitz}, {Hemar},
  {Ibbetson}, {Mashal}, {Meistas}, {Dziembowski}, {Pamyatnykh}, {Moskalik},
  {Zola}, {Pajdosz}, {Krzesinski}, {Solheim}, {Bard}, {Massacand}, {Breger},
  {Gelbmann}, {Paunzen}, \& {North}}]{1997MNRAS.286..303H}
{Handler}, G., {Pikall}, H., {O'Donoghue}, D., {et~al.} 1997, \mnras, 286, 303

\bibitem[{{Hermes} {et~al.}(2014){Hermes}, {G{\"a}nsicke}, {Koester}, {Bours},
  {Townsley}, {Farihi}, {Marsh}, {Littlefair}, {Dhillon}, {Gianninas},
  {Breedt}, \& {Raddi}}]{2014MNRAS.444.1674H}
{Hermes}, J.~J., {G{\"a}nsicke}, B.~T., {Koester}, D., {et~al.} 2014, \mnras,
  444, 1674

\bibitem[{{Hermes} {et~al.}(2013{\natexlab{a}}){Hermes}, {Montgomery},
  {Gianninas}, {Winget}, {Brown}, {Harrold}, {Bell}, {Kenyon}, {Kilic}, \&
  {Castanheira}}]{2013MNRAS.436.3573H}
{Hermes}, J.~J., {Montgomery}, M.~H., {Gianninas}, A., {et~al.}
  2013{\natexlab{a}}, \mnras, 436, 3573

\bibitem[{{Hermes} {et~al.}(2013{\natexlab{b}}){Hermes}, {Montgomery},
  {Winget}, {Brown}, {Gianninas}, {Kilic}, {Kenyon}, {Bell}, \&
  {Harrold}}]{2013ApJ...765..102H}
{Hermes}, J.~J., {Montgomery}, M.~H., {Winget}, D.~E., {et~al.}
  2013{\natexlab{b}}, \apj, 765, 102

\bibitem[{{Hermes} {et~al.}(2012){Hermes}, {Montgomery}, {Winget}, {Brown},
  {Kilic}, \& {Kenyon}}]{2012ApJ...750L..28H}
{Hermes}, J.~J., {Montgomery}, M.~H., {Winget}, D.~E., {et~al.} 2012, \apjl,
  750, L28

\bibitem[{{Iglesias} \& {Rogers}(1996)}]{1996ApJ...464..943I}
{Iglesias}, C.~A. \& {Rogers}, F.~J. 1996, \apj, 464, 943

\bibitem[{{Istrate}(2015)}]{2015ASPC..493..487I}
{Istrate}, A.~G. 2015, in Astronomical Society of the Pacific Conference
  Series, Vol. 493, 19th European Workshop on White Dwarfs, ed. P.~{Dufour},
  P.~{Bergeron}, \& G.~{Fontaine}, 487

\bibitem[{{Istrate} {et~al.}(2016{\natexlab{a}}){Istrate}, {Fontaine},
  {Gianninas}, {Grassitelli}, {Marchant}, {Tauris}, \&
  {Langer}}]{2016A&A...595L..12I}
{Istrate}, A.~G., {Fontaine}, G., {Gianninas}, A., {et~al.} 2016{\natexlab{a}},
  \aap, 595, L12

\bibitem[{{Istrate} {et~al.}(2016{\natexlab{b}}){Istrate}, {Marchant},
  {Tauris}, {Langer}, {Stancliffe}, \& {Grassitelli}}]{2016A&A...595A..35I}
{Istrate}, A.~G., {Marchant}, P., {Tauris}, T.~M., {et~al.} 2016{\natexlab{b}},
  \aap, 595, A35

\bibitem[{{Itoh} {et~al.}(1996){Itoh}, {Hayashi}, {Nishikawa}, \&
  {Kohyama}}]{1996ApJS..102..411I}
{Itoh}, N., {Hayashi}, H., {Nishikawa}, A., \& {Kohyama}, Y. 1996, \apjs, 102,
  411

\bibitem[{{Kawaler}(1988)}]{1988IAUS..123..329K}
{Kawaler}, S.~D. 1988, in IAU Symposium, Vol. 123, Advances in Helio- and
  Asteroseismology, ed. J.~{Christensen-Dalsgaard} \& S.~{Frandsen}, 329

\bibitem[{{Kepler} {et~al.}(2014){Kepler}, {Fraga}, {Winget}, {Bell},
  {C{\'o}rsico}, \& {Werner}}]{2014MNRAS.442.2278K}
{Kepler}, S.~O., {Fraga}, L., {Winget}, D.~E., {et~al.} 2014, \mnras, 442, 2278

\bibitem[{{Kepler} {et~al.}(2012){Kepler}, {Pelisoli}, {Pe{\c c}anha}, {Costa},
  {Fraga}, {Hermes}, {Winget}, {Castanheira}, {C{\'o}rsico}, {Romero},
  {Althaus}, {Kleinman}, {Nitta}, {Koester}, {K{\"u}lebi}, {Jordan}, \&
  {Kanaan}}]{2012ApJ...757..177K}
{Kepler}, S.~O., {Pelisoli}, I., {Pe{\c c}anha}, V., {et~al.} 2012, \apj, 757,
  177

\bibitem[{{Kilic} {et~al.}(2011){Kilic}, {Brown}, {Allende Prieto},
  {Ag{\"u}eros}, {Heinke}, \& {Kenyon}}]{2011ApJ...727....3K}
{Kilic}, M., {Brown}, W.~R., {Allende Prieto}, C., {et~al.} 2011, \apj, 727, 3

\bibitem[{{Kilic} {et~al.}(2012){Kilic}, {Brown}, {Allende Prieto}, {Kenyon},
  {Heinke}, {Ag{\"u}eros}, \& {Kleinman}}]{2012ApJ...751..141K}
{Kilic}, M., {Brown}, W.~R., {Allende Prieto}, C., {et~al.} 2012, \apj, 751,
  141

\bibitem[{{Kilic} {et~al.}(2015){Kilic}, {Hermes}, {Gianninas}, \&
  {Brown}}]{2015MNRAS.446L..26K}
{Kilic}, M., {Hermes}, J.~J., {Gianninas}, A., \& {Brown}, W.~R. 2015, \mnras,
  446, L26

\bibitem[{{Koester} {et~al.}(2009){Koester}, {Voss}, {Napiwotzki},
  {Christlieb}, {Homeier}, {Lisker}, {Reimers}, \&
  {Heber}}]{2009A&A...505..441K}
{Koester}, D., {Voss}, B., {Napiwotzki}, R., {et~al.} 2009, \aap, 505, 441

\bibitem[{{Magni} \& {Mazzitelli}(1979)}]{1979A&A....72..134M}
{Magni}, G. \& {Mazzitelli}, I. 1979, \aap, 72, 134

\bibitem[{{Maxted} {et~al.}(2011){Maxted}, {Anderson}, {Burleigh}, {Collier
  Cameron}, {Heber}, {G{\"a}nsicke}, {Geier}, {Kupfer}, {Marsh}, {Nelemans},
  {O'Toole}, {{\O}stensen}, {Smalley}, \& {West}}]{2011MNRAS.418.1156M}
{Maxted}, P.~F.~L., {Anderson}, D.~R., {Burleigh}, M.~R., {et~al.} 2011,
  \mnras, 418, 1156

\bibitem[{{Maxted} {et~al.}(2014){Maxted}, {Serenelli}, {Marsh}, {Catal{\'a}n},
  {Mahtani}, \& {Dhillon}}]{2014MNRAS.444..208M}
{Maxted}, P.~F.~L., {Serenelli}, A.~M., {Marsh}, T.~R., {et~al.} 2014, \mnras,
  444, 208

\bibitem[{{Maxted} {et~al.}(2013){Maxted}, {Serenelli}, {Miglio}, {Marsh},
  {Heber}, {Dhillon}, {Littlefair}, {Copperwheat}, {Smalley}, {Breedt}, \&
  {Schaffenroth}}]{2013Natur.498..463M}
{Maxted}, P.~F.~L., {Serenelli}, A.~M., {Miglio}, A., {et~al.} 2013, \nat, 498,
  463

\bibitem[{{Nelson} {et~al.}(2004){Nelson}, {Dubeau}, \&
  {MacCannell}}]{2004ApJ...616.1124N}
{Nelson}, L.~A., {Dubeau}, E., \& {MacCannell}, K.~A. 2004, \apj, 616, 1124

\bibitem[{{O'Donoghue}(1994)}]{1994MNRAS.270..222O}
{O'Donoghue}, D. 1994, \mnras, 270, 222

\bibitem[{{Papar{\'o}} {et~al.}(2013){Papar{\'o}}, {Bogn{\'a}r}, {Plachy},
  {Moln{\'a}r}, \& {Bradley}}]{2013MNRAS.432..598P}
{Papar{\'o}}, M., {Bogn{\'a}r}, Z., {Plachy}, E., {Moln{\'a}r}, L., \&
  {Bradley}, P.~A. 2013, \mnras, 432, 598

\bibitem[{{Pech} \& {Vauclair}(2006)}]{2006A&A...453..219P}
{Pech}, D. \& {Vauclair}, G. 2006, \aap, 453, 219

\bibitem[{{Pech} {et~al.}(2006){Pech}, {Vauclair}, \&
  {Dolez}}]{2006A&A...446..223P}
{Pech}, D., {Vauclair}, G., \& {Dolez}, N. 2006, \aap, 446, 223

\bibitem[{{Romero} {et~al.}(2012){Romero}, {C{\'o}rsico}, {Althaus}, {Kepler},
  {Castanheira}, \& {Miller Bertolami}}]{2012MNRAS.420.1462R}
{Romero}, A.~D., {C{\'o}rsico}, A.~H., {Althaus}, L.~G., {et~al.} 2012, \mnras,
  420, 1462

\bibitem[{{Romero} {et~al.}(2013){Romero}, {Kepler}, {C{\'o}rsico}, {Althaus},
  \& {Fraga}}]{2013ApJ...779...58R}
{Romero}, A.~D., {Kepler}, S.~O., {C{\'o}rsico}, A.~H., {Althaus}, L.~G., \&
  {Fraga}, L. 2013, \apj, 779, 58

\bibitem[{{Saio} {et~al.}(1983){Saio}, {Winget}, \&
  {Robinson}}]{1983ApJ...265..982S}
{Saio}, H., {Winget}, D.~E., \& {Robinson}, E.~L. 1983, \apj, 265, 982

\bibitem[{{Serenelli} {et~al.}(2002){Serenelli}, {Althaus}, {Rohrmann}, \&
  {Benvenuto}}]{2002MNRAS.337.1091S}
{Serenelli}, A.~M., {Althaus}, L.~G., {Rohrmann}, R.~D., \& {Benvenuto}, O.~G.
  2002, \mnras, 337, 1091

\bibitem[{{Steinfadt} {et~al.}(2010){Steinfadt}, {Bildsten}, \&
  {Arras}}]{2010ApJ...718..441S}
{Steinfadt}, J.~D.~R., {Bildsten}, L., \& {Arras}, P. 2010, \apj, 718, 441

\bibitem[{{Tassoul}(1980)}]{1980ApJS...43..469T}
{Tassoul}, M. 1980, \apjs, 43, 469

\bibitem[{{Tassoul} {et~al.}(1990){Tassoul}, {Fontaine}, \&
  {Winget}}]{1990ApJS...72..335T}
{Tassoul}, M., {Fontaine}, G., \& {Winget}, D.~E. 1990, \apjs, 72, 335

\bibitem[{{Tremblay} {et~al.}(2015){Tremblay}, {Gianninas}, {Kilic}, {Ludwig},
  {Steffen}, {Freytag}, \& {Hermes}}]{2015ApJ...809..148T}
{Tremblay}, P.-E., {Gianninas}, A., {Kilic}, M., {et~al.} 2015, \apj, 809, 148

\bibitem[{{Unno} {et~al.}(1989){Unno}, {Osaki}, {Ando}, {Saio}, \&
  {Shibahashi}}]{1989nos..book.....U}
{Unno}, W., {Osaki}, Y., {Ando}, H., {Saio}, H., \& {Shibahashi}, H. 1989,
  {Nonradial oscillations of stars}, ed. T.~University~of Tokyo~Press

\bibitem[{{Van Grootel} {et~al.}(2013){Van Grootel}, {Fontaine}, {Brassard}, \&
  {Dupret}}]{2013ApJ...762...57V}
{Van Grootel}, V., {Fontaine}, G., {Brassard}, P., \& {Dupret}, M.-A. 2013,
  \apj, 762, 57

\bibitem[{{Winget} \& {Kepler}(2008)}]{2008ARA&A..46..157W}
{Winget}, D.~E. \& {Kepler}, S.~O. 2008, \araa, 46, 157

\bibitem[{{Zhang} {et~al.}(2016){Zhang}, {Fu}, {Li}, {Ren}, \&
  {Luo}}]{2016ApJ...821L..32Z}
{Zhang}, X.~B., {Fu}, J.~N., {Li}, Y., {Ren}, A.~B., \& {Luo}, C.~Q. 2016,
  \apjl, 821, L32

\end{thebibliography}

\end{document}